\RequirePackage{fix-cm} 
\documentclass[a4paper, twoside, reqno, dvips, 12pt]{amsart}
\usepackage{fixltx2e}   

\usepackage{etex}

\usepackage[latin1]{inputenc}
\usepackage[T1]{fontenc}

\usepackage{eucal}
\usepackage{esint}
\usepackage{dsfont}
\usepackage{xspace}
\usepackage{amsgen}
\usepackage{amsthm}
\usepackage{amssymb}
\usepackage{amsmath}
\usepackage{upgreek}
\usepackage{wasysym}
\usepackage{amsfonts}
\usepackage{stmaryrd}
\usepackage{mathtools}

\usepackage{mathrsfs}
\DeclareMathAlphabet{\mathscrbf}{OMS}{mdugm}{b}{n}

\usepackage{a4wide}

\usepackage[dvipsnames, table]{xcolor}
\definecolor{bckg}{RGB}{20.8, 20.8, 20.8}
\definecolor{oneblue}{rgb}{0.0, 0.0, 0.85}
\definecolor{Lightblue}{RGB}{214, 214, 214}
\definecolor{bluepigment}{rgb}{0.2, 0.2, 0.6}
\definecolor{charcoal}{rgb}{0.21, 0.27, 0.31}
\definecolor{denimblue}{rgb}{0.08, 0.38, 0.74}
\definecolor{Lightgray}{rgb}{0.89, 0.89, 0.89}
\definecolor{darkgrey}{rgb}{0.273, 0.281, 0.30}
\definecolor{darkelectricblue}{rgb}{0.33, 0.41, 0.47}

\usepackage{psfrag}
\usepackage{graphicx}
\usepackage{subfigure}
\usepackage{morefloats}
\usepackage{indentfirst}

\usepackage{acronym}
\usepackage{microtype}
\usepackage[labelsep=period,%
            labelfont={bf,sf,color=bluepigment},%
            justification=raggedright]{caption}

\usepackage[perpage, symbol]{footmisc}

\usepackage[usenames, dvipsnames]{pstricks}
\usepackage{epsfig}
\usepackage{pst-grad} 
\usepackage{pst-plot} 

\usepackage[colorlinks,
          urlcolor=oneblue,
          linkcolor=denimblue,
          citecolor=NavyBlue,
          bookmarksopen=false,
          pdfpagemode=UseNone,
          pagebackref]{hyperref}

\usepackage[sort&compress, comma, square, numbers]{natbib}

\usepackage[explicit]{titlesec}

\titleformat{\section}
  {\color{NavyBlue}\Large\sffamily\bfseries}
  {}
  {0em}
  {\colorbox{bckg!5}{\parbox{\dimexpr\linewidth-2\fboxsep\relax}{\centering\thesection. #1}}}
  [\vspace*{0.33em}]

\titleformat{name=\section,numberless}
  {\color{NavyBlue}\Large\sffamily\bfseries}
  {}
  {0.0em}
  {\colorbox{bckg!10}{\parbox{\dimexpr\linewidth-2\fboxsep\relax}{\centering#1}}}
  [\vspace*{0.33em}]

\titleformat{\subsection}
  {\color{NavyBlue}\large\sffamily\bfseries}
  {}
  {0.0em}
  {\colorbox{bckg!5}{\parbox{\dimexpr\linewidth-2\fboxsep\relax}{\centering\thesubsection. #1}}}
  [\vspace*{0.33em}]

\titleformat{name=\subsection,numberless}
  {\color{NavyBlue}\Large\sffamily\bfseries}
  {}
  {0em}
  {\colorbox{bckg!10}{\parbox{\dimexpr\linewidth-2\fboxsep\relax}{\centering#1}}}
  [\vspace*{0.33em}]

\titleformat{\subsubsection}
  {\color{bluepigment}\sffamily\normalsize\bfseries}
  {\thesubsubsection}
  {0.5em}
  {#1}
  [\vspace*{0.33em}]

\titleformat{\paragraph}[runin]
  {\color{bluepigment}\sffamily\small\bfseries}
  {}
  {0em}
  {#1}

\titlespacing{\section}{1.0em}{1.5em plus 2pt minus 2pt}%
{1.0em plus 2pt minus 2pt}[0em]
\titlespacing{\subsection}{1.0em}{1.5em plus 2pt minus 2pt}%
{1.0em}[0em]
\titlespacing{\subsubsection}{1.0em}{1.5em plus 2pt minus 2pt}%
{1.0em plus 2pt minus 2pt}[0em]

\usepackage{titletoc}

\contentsmargin{0.5em}
\newlength{\tocsep} 
\titlecontents{section}[\tocsep]
  {\addvspace{10pt}\bfseries\sffamily}
  {\contentslabel[\thecontentslabel]{\tocsep}}
  {}
  {\ \titlerule*[0.75pc]{.}\ \thecontentspage}
  []
\titlecontents{subsection}[\tocsep]
  {\addvspace{8pt}\sffamily}
  {\contentslabel[\thecontentslabel]{\tocsep}}
  {}
  {\ \titlerule*[0.5pc]{.}\ \thecontentspage}
  []
\titlecontents*{subsubsection}[\tocsep]
  {\addvspace{2pt}\footnotesize\sffamily}
  {}
  {}
  {\ \titlerule*[0.35pc]{.}\ \thecontentspage}
  [\\*]

\makeatletter
\def\@setauthors{%
  \begingroup
  \def\thanks{\protect\thanks@warning}%
  \trivlist
  \centering\footnotesize \@topsep30\p@\relax
  \advance\@topsep by -\baselineskip
  \item\relax
  \author@andify\authors
  \def\\{\protect\linebreak}%
  \textsc{\normalsize\textcolor{darkelectricblue}{\authors}}%
  \ifx\@empty\contribs
  \else
    ,\penalty-3 \space \@setcontribs
    \@closetoccontribs
  \fi
  \endtrivlist
  \endgroup
}
\def\@settitle{\begin{center}%
  \baselineskip14\p@\relax
    \bfseries
    \textsc{\Large\textcolor{charcoal}{\@title}}
  \end{center}%
}
\makeatother

\usepackage{enumitem}
\setlist[description]{%
  topsep=30pt,               
  itemsep=5pt,               
  font={\bfseries\sffamily\color{NavyBlue}}, 
}

\usepackage{fancyhdr}
\usepackage{lastpage}

\newcommand*\Title{\textcolor{bluepigment}{Accurate numerical simulation of moisture front}}
\newcommand*\Authors{\textcolor{bluepigment}{J.~Berger, S.~Gasparin, D.~Dutykh \& N.~Mendes}}
\newcommand*{\plogo}{\textcolor{gray}{{\texttt{arXiv.org} / \textsc{hal}}}} 

\numberwithin{equation}{section}

\newcommand{\CN}{\textsc{Crank}--\textsc{Nicolson}}
\newcommand{\mCN}{modified \textsc{Crank}--\textsc{Nicolson}}
\newcommand{\SG}{\textsc{Scharfetter}--\textsc{Gummel}}

\newcommand*\egal{\ = \ }
\newcommand*\plus{\ + \ }
\newcommand*\moins{\ - \ }

\renewcommand{\O}{\mathcal{O}}
\newcommand*{\Ox}{\Omega_{\, x}}

\newcommand{\f}{\mathrm{f}}
\newcommand{\const}{\mathrm{const}}

\newcommand{\Bm}{\mathcal{B}\biggl( \, -\ \frac{a \ \Delta x}{\nu} \, \biggr)}
\newcommand{\Bp}{\mathcal{B}\biggl( \, \frac{a \ \Delta x}{\nu} \, \biggr)}

\newcommand{\bm}{\mathrm{b}^{-}}
\newcommand{\bp}{\mathrm{b}^{+}}
\newcommand{\bpm}{\mathrm{b}^{\pm}}
\newcommand{\BivL}{\mathrm{Bi}_{\,v}^{\,\mathrm{L}}}
\newcommand{\BivR}{\mathrm{Bi}_{\,v}^{\,\mathrm{R}}}
\newcommand{\cm}{c_{\,m}}
\newcommand{\cms}{c_{\,m}^{\,\star}}
\newcommand{\dm}{d_{\,m}}
\newcommand{\dmref}{d_{\,m}^{\,0}}
\newcommand{\dms}{d_{\,m}^{\,\star}}

\newcommand{\glL}{g_{\,l}^{\,\mathrm{L}}}

\newcommand{\glsL}{g_{\,l}^{\,\star \mathrm{L}}}

\newcommand{\hvL}{h_{\,v}^{\,\mathrm{L}}}
\newcommand{\hvR}{h_{\,v}^{\,\mathrm{R}}}
\newcommand{\kl}{k_{\,l}}
\newcommand{\kv}{k_{\,v}}
\newcommand{\Pc}{P_{\,c}}
\newcommand{\Pe}{\mathrm{P\mbox{\'e}}}
\newcommand{\Ps}{P_{\,s}}
\newcommand{\Pv}{P_{\,v}}
\newcommand{\Pvi}{P_{\,v}^{\,i}}
\newcommand{\PvL}{P_{\,v}^{\,\mathrm{L}}}
\newcommand{\PvR}{P_{\,v}^{\,\mathrm{R}}}
\newcommand{\Rv}{R_{\,v}}
\newcommand{\tref}{t^{\,0}}
\newcommand{\uL}{u^{\,\mathrm{L}}}
\newcommand{\uR}{u^{\,\mathrm{R}}}
\newcommand{\rholv}{\rho_{\,l+v}}
\newcommand{\xs}{x^{\,\star}}
\newcommand{\ts}{t^{\,\star}}

\newcommand{\dt}{\Delta t}
\newcommand{\dx}{\Delta x}

\newcommand*\pd[2]{\frac{\partial #1}{\partial #2}}

\newcommand{\eqdef}{\mathop{\stackrel{\,\mathrm{def}}{:=}\,}}

\newcommand{\sign}{\mathop{\mathrm{sign}}}

\newcommand{\half}{{\textstyle{1\over2}}}

\newcommand{\dix}[1]{ \cdot 10^{\,#1}}

\def\R{\mbox{I\hspace{-.15em}R}}

\newcommand{\PE}{\mathrm{Pe}}

\newcommand{\unite}[1]{ $\mathsf{#1 \,}$}

\begin{document}

\title[\Title]{Accurate numerical simulation of moisture front in porous material}

\author[J.~Berger]{Julien Berger$^*$}
\address{\textbf{J.~Berger:} Thermal Systems Laboratory, Mechanical Engineering Graduate Program, Pontifical Catholic University of Paran\'a, Rua Imaculada Concei\c{c}\~{a}o, 1155, CEP: 80215-901, Curitiba -- Paran\'a, Brazil}
\email{Julien.Berger@pucpr.edu.br}
\urladdr{https://www.researchgate.net/profile/Julien\_Berger3/}
\thanks{$^*$ Corresponding author}

\author[S.~Gasparin]{Suelen Gasparin}
\address{\textbf{S.~Gasparin:} Thermal Systems Laboratory, Mechanical Engineering Graduate Program, Pontifical Catholic University of Paran\'a, Rua Imaculada Concei\c{c}\~{a}o, 1155, CEP: 80215-901, Curitiba -- Paran\'a, Brazil}
\email{suelengasparin@hotmail.com}
\urladdr{https://www.researchgate.net/profile/Suelen\_Gasparin/}

\author[D.~Dutykh]{Denys Dutykh}
\address{\textbf{D.~Dutykh:} LAMA, UMR 5127 CNRS, Universit\'e Savoie Mont Blanc, Campus Scientifique, F-73376 Le Bourget-du-Lac Cedex, France}
\email{Denys.Dutykh@univ-savoie.fr}
\urladdr{http://www.denys-dutykh.com/}

\author[N.~Mendes]{Nathan Mendes}
\address{\textbf{N.~Mendes:} Thermal Systems Laboratory, Mechanical Engineering Graduate Program, Pontifical Catholic University of Paran\'a, Rua Imaculada Concei\c{c}\~{a}o, 1155, CEP: 80215-901, Curitiba -- Paran\'a, Brazil}
\email{Nathan.Mendes@pucpr.edu.br}
\urladdr{https://www.researchgate.net/profile/Nathan\_Mendes/}



\begin{titlepage}
\thispagestyle{empty} 
\noindent
{\Large Julien \textsc{Berger}}\\
{\it\textcolor{gray}{Pontifical Catholic University of Paran\'a, Brazil}}
\\[0.02\textheight]
{\Large Suelen \textsc{Gasparin}}\\
{\it\textcolor{gray}{Pontifical Catholic University of Paran\'a, Brazil}}
\\[0.02\textheight]
{\Large Denys \textsc{Dutykh}}\\
{\it\textcolor{gray}{CNRS, Universit\'e Savoie Mont Blanc, France}}
\\[0.02\textheight]
{\Large Nathan \textsc{Mendes}}\\
{\it\textcolor{gray}{Pontifical Catholic University of Paran\'a, Brazil}}
\\[0.10\textheight]

\colorbox{Lightblue}{
  \parbox[t]{1.0\textwidth}{
    \centering\huge\sc
    \vspace*{0.7cm}
    
    \textcolor{bluepigment}{Accurate numerical simulation of moisture front in porous material}

    \vspace*{0.7cm}
  }
}

\vfill 

\raggedleft     
{\large \plogo} 
\end{titlepage}


\newpage
\thispagestyle{empty} 
\par\vspace*{\fill}   
\begin{flushright} 
{\textcolor{denimblue}{\textsc{Last modified:}} \today}
\end{flushright}


\newpage
\maketitle
\thispagestyle{empty}


\begin{abstract}

When comparing measurements to numerical simulations of moisture transfer through porous materials a rush of the experimental moisture front is commonly observed in several works shown in the literature, with transient models that consider only the diffusion process. Thus, to overcome the discrepancies between the experimental and the numerical models, this paper proposes to include the moisture advection transfer in the governing equation. To solve the advection-diffusion or the so-called convection differential equation, it is first proposed two efficient numerical schemes and their efficiencies are investigated for both linear and nonlinear cases. The first scheme, \textsc{Scharfetter}--\textsc{Gummel} (SG), presents a \textsc{Courant}--\textsc{Friedrichs}--\textsc{Lewy} (CFL) condition but is more accurate and faster than the second scheme, the well-known \textsc{Crank}--\textsc{Nicolson} approach. Furthermore, the SG scheme has the advantages of being well-balanced and asymptotically preserved. Then, to conclude, results of the convective moisture transfer problem obtained with the SG numerical scheme are compared to experimental data from the literature. The inclusion of an advective term in the model may clearly lead to better results than purely diffusive models.

\bigskip
\noindent \textbf{\keywordsname:} Advection-diffusion equation; numerical methods; benchmarking experimental data; \SG ~scheme; convective moisture transport; hygroscopic materials \\


\end{abstract}


\newpage
\tableofcontents
\thispagestyle{empty}


\newpage
\section{Introduction}

Moisture in porous building elements can affect indoor air quality, thermal comfort and energy consumption/demand. excessive level of moisture may also damage the construction quality and their durability and can lead to mould growth on the inside surface \cite{Harris2001, Berger2015a}.

In order to predict those effects in buildings, moisture transfer models have been integrated in early nineties in simulation tools such as \texttt{Delphin} \cite{BauklimatikDresden2011}, \texttt{MATCH} \cite{Rode2003}, \texttt{MOIST} \cite{Burch1993}, \texttt{WUFI} \cite{IBP2005}, \texttt{Umidus} \cite{Mendes1997, Mendes1999} and \texttt{Blast} \cite{Liesen1994}. In the frame of the International Energy Agency Annex~$41\,$, detailed models and their successful applications for accurate assessment of hygrothermal transfer in buildings have been reported \cite{Woloszyn2008}.


\subsection{Problem statement}

\begin{figure}
\centering
\def\svgwidth{1\textwidth}
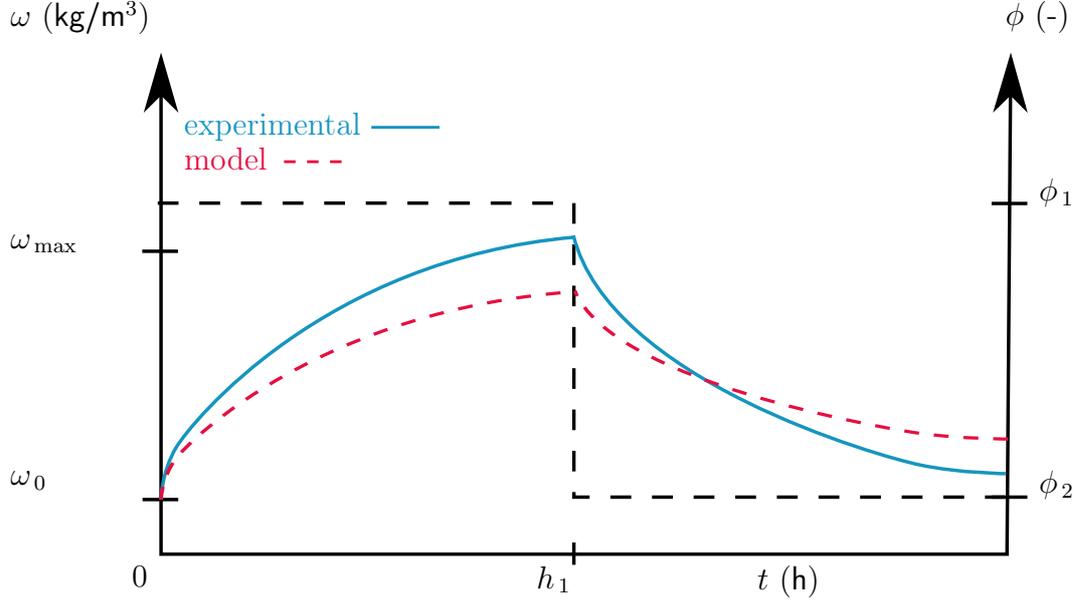
\caption{\small\em Illustration of the discrepancies observed when comparing experimental data to results from numerical model of moisture transfer in porous material.}
\label{intro_fig:err_mod_exp}
\end{figure}

Nevertheless, some discrepancies normally appear when comparing the results from numerical models and experimental data, as illustrated in Figure~\ref{intro_fig:err_mod_exp}. A material, with an initial moisture content $w_{\,0}$, is submitted to an adsorption phase at $\phi_{\,1}$ and then to a desorption phase at  $\phi_{\,2} \,$. Results of the simulation under estimate the adsorption process or over estimate the desorption process. In other terms, the experimental moisture front always rushes faster than the simulation results. Numerous studies state similar observations.

In \cite{McClung2014} four cross-laminated timber wall assemblies were studied monitoring a test wall during one year period. The boundary conditions corresponded to outside weather and fixed in time for indoor side. The panels were initially wetted and their drying behaviour were analysed. Simulations were performed with the \texttt{WUFI} program (based on \textsc{Kunzel} diffusion model \cite{Kunzel1995}), using material properties based on laboratory characterisation.

In \cite{Talukdar2007, Talukdar2007a}, spruce plywood and cellulose insulation were evaluated considering single-step change increase in humidity or adsorption and desorption cycle tests. The model used to compare the experimental data is based on moisture diffusion due to water vapour density or total gas pressure difference.

In \cite{Samri2008}, autoclaved and hemp concretes are used combined with various experimental designs. Data are compared with the \textsc{Kunzel} diffusion model \cite{Kunzel1995}. The comparison reveals the same type of discrepancies, specially for the design operating four $24$ hour steps of temperature and relative humidity.

In \cite{James2010}, gypsum boards were conditioned to adsorption and desorption cycles of $(30\%-70\%-30\%)$ relative humidity. The whole experiment was conducted for $48\mathsf{h}$ under isothermal conditions. The numerical results, in terms of relative humidity, obtained with models from eight different institutions, were compared to experimental data. All the models predicted transient behaviour slower than experimental data.

In \cite{Lelievre2014, Colinart2016} experiments were performed in a climatic chamber with hemp concrete samples. Slow and fast cycling tests of adsorption and desorption were done. The discrepancies between experimental data and model results were reduced by considering the hysteresis of the material moisture capacity. In \cite{Fabbri2015}, other hygrothermal data were provided for hemp concrete and compared to a numerical model without hysteresis effect considerations. The conclusions underlined the good tendencies but nevertheless with some lack of accuracies. In \cite{Maalouf2014}, experiments for similar materials under climatic variations were performed. Influence of material properties and convective coefficients were investigated to reduce the discrepancies with experimental data.

Some experimental designs were also operated at the building scale. In \cite{Labat2015, Piot2011}, a wooden-frame house was instrumented. Vapour was generated during certain periods. The comparison with the numerical model was during and after those periods and some discrepancies were observed in the transient behaviour. An experimental benchmark is presented in \cite{VanBelleghem2011}, using calcium silicate boards submitted to five adsorption and desorption cycles $(50\%-70\%-50\%)$. The model used for comparison included two sub-models considering coupled heat and moisture transfer equations in the material and in the air within the climatic chamber.

All those studies highlighted slower transient behaviour of the results obtained by numerical models comparing to experimental data. The observations are particularly valuable for hygroscopic materials. The models are based on the coupled heat and moisture diffusion in porous materials.


\subsection{Objectives of the paper}

Some attempts have been done to reduce those discrepancies. Among others, in \cite{Kwiatkowski2009, Lelievre2014, Colinart2016}, the hysteresis of the sorption material capacity was considered. In \cite{Olek2016} a non-\textsc{Fickian} moisture diffusion model was proposed for thermally modified wood. A possible explanation of the slower transient behaviour of the numerical results is the absence of advection transfer in the proposed model. When the advective and diffusive fluxes have the same direction, the advection mechanism increases the moisture front velocity. Hygroscopics materials such as wood fibre board, gypsum board and aerated cellular concreted have a larger air permeability, almost three orders of magnitude higher, if compared with the concrete one \cite{KumarKumaran1996}. Some numerical models have been developed considering moisture advection \cite{Belleudy2016}. However, to the knowledge of the authors, no comparison with experimental studies of adsorption/desorption cycles for building materials have been accomplished.

Thus, the objectives of this paper are basically two. First, it aims at analysing the numerical schemes to solve an advective-diffusive problem or the so-called convective moisture transfer in porous materials, represented by a model proposed in Section~\ref{sec:Moisture_convection}. After a brief recall of the fundamentals and objectives of numerical methods, the \CN ~and the \SG ~schemes are then described. The primer has been extensively used to solve advective-diffusive equation as for instance in \cite{Perrochet1993}. The latter is a relatively innovative approach, despite being firstly proposed in 1969, and presents several advantages that will be discussed for both linear and nonlinear problems. Then, the second objective is to illustrate the influence of the moisture convection hypothesis on the comparison with the experimental results. Thus, in the last section, the results of the \SG ~scheme are compared to an isothermal experiment from \cite{James2010}.

\section{Moisture transfer in porous materials by diffusion and advection}
\label{sec:Moisture_convection}

The physical problem involves one-dimension moisture convection through a porous material defined by the spatial domain $\Ox \egal [\, 0, \, L \,] $. The moisture transfer occurs due to capillary migration, vapour diffusion and advection of the vapour phase. The physical problem can be formulated as \cite{Janssen2014, Tariku2010, Belleudy2016}:
\begin{align}\label{eq:moisture_equation_1D}
& \pd{\rholv}{t} \egal \pd{}{x} \left( \, \kl \, \pd{\Pc}{x} \plus \kv \, \pd{\Pv}{x} \, \right) 
\moins \pd{}{x}\left( \, \frac{\Pv}{\Rv \ T} \, v \, \right) \,,
\end{align}
where $\rholv$ is the volumetric moisture content of the material, $\kv$ and $\kl$, the vapour and liquid permeabilities, $\Pv$, the vapour pressure, $T$, the temperature, $v$, the mass average velocity and, $\Rv$, the water vapour gas constant. Eq.~\eqref{eq:moisture_equation_1D} can be written using the vapour pressure $\Pv$ as the driving potential. For this, we consider the physical relation, known as the \textsc{Kelvin} equation, between $\Pv$ and $\Pc \,$:
\begin{align*}
\Pc & \egal \rho_{\,l} \, \Rv \, T \, \ln\left(\frac{\Pv}{\Ps(T)}\right) \,,\\
\pd{\Pc}{\Pv} & \egal \rho_{\,l} \, \frac{R_{\,v} \, T}{\Pv} \,.
\end{align*}
Thus we have:
\begin{align*}
\pd{\Pc}{x} \egal \pd{\Pc}{\Pv} \, \pd{\Pv}{x} \plus \pd{\Pc}{T} \, \pd{T}{x} \,.
\end{align*}
The temperature remains the same at the boundaries. Even if heat transfer occurs in the material due to latent heat evaporation, the temperature variations in the material are assumed negligible. Thus, the second right-hand term vanishes and we obtain:
\begin{align*}
& \pd{\Pc}{x} \egal  \rho_{\,l} \, \frac{\Rv \, T}{\Pv} \, \pd{\Pv}{x} \,.
\end{align*}
In addition, we have:
\begin{align*}
& \pd{\rholv}{t} \egal \pd{\rholv}{\phi} \, \pd{\phi}{\Pv} \, \pd{\Pv}{t} \plus \pd{\rholv}{T} \, \pd{T}{t} \,.
\end{align*}
Under isothermal conditions, the second right-hand term of the equation above also vanishes. Considering the relation $\rholv \egal \f(\phi) \egal \f(\Pv,T)$, obtained from material properties and from the relation between the vapour pressure $\Pv$ and the relative humidity $\phi$, we get: 
\begin{align*}
& \pd{\rholv}{t} \egal \f^{\,\prime}(\Pv) \; \frac{1}{\Ps} \; \pd{\Pv}{t} \,.
\end{align*}
For the advection term of Eq.~\eqref{eq:moisture_equation_1D}, with the assumption of isothermal conditions and constant mass average velocity $\mathsf{v}$, we can write: 
\begin{align*}
& \pd{}{x}\left( \, \frac{\Pv}{\Rv \ T} \, \mathsf{v} \, \right) \simeq \frac{\mathsf{v}}{\Rv \ T} \, \pd{\Pv}{x} \,.
\end{align*}
Eq.~\eqref{eq:moisture_equation_1D} can be therefore rewritten as:
\begin{align}\label{eq:moisture_equation_1D_v2}
& \f^{\,\prime}(\Pv) \; \frac{1}{\Ps} \; \pd{\Pv}{t} \egal \pd{}{x} \Biggl[ \, \Bigl( \, \kl \, \frac{\rho_{\,l} \, \Rv \, T}{\Pv} \plus \kv \, \Bigr) \, \pd{\Pv}{x} \, \Biggr] \moins \frac{v}{\Rv \ T} \, \pd{\Pv}{x} \,.
\end{align}
The material properties $\f^{\,\prime}(\Pv)$, $\kl$ and $\kv$ depend on the vapour pressure $\Pv \,$. We denote $\dm \egal \kl \, \dfrac{\rho_{\,l} \, \Rv \, T}{\Pv} \plus \kv $ the global moisture transport coefficient and $\cm \egal \f^{\,\prime}(\Pv) \; \dfrac{1}{\Ps}$ the moisture storage coefficient. Thus, Eq.~\eqref{eq:moisture_equation_1D_v2} becomes:
\begin{align}\label{eq:moisture_equation_1D_v3}
  & \cm \; \pd{\Pv}{t} \egal \pd{}{x} \Biggl[ \, \dm \, \pd{\Pv}{x} \, \Biggr] \moins \frac{\mathsf{v}}{\Rv \ T} \, \pd{\Pv}{x} \,.
\end{align}
At the material bounding surfaces, \textsc{Robin}-type boundary conditions are considered:
\begin{align}\label{eq:bc}
\dm \, \pd{\Pv}{x} &\egal 
\hvL \cdot \left( \, \Pv \moins \PvL \, \right) \moins \glL \, , && x \egal 0 \,, \\
 - \dm \, \pd{\Pv}{x} &\egal 
\hvR \cdot \left( \, \Pv \moins \PvR \, \right)\, ,&& x \egal L \,,
\end{align}
where $\PvL$ and $\PvR$ are the vapour pressure of the ambient air and $\glL$ is the liquid flow (driving rain) at the left bounding surface. We consider a uniform vapour pressure distribution as initial condition:
\begin{align}\label{eq:ic}
 \Pv &\egal \Pvi \,, && t\egal0 \,.
\end{align}

While performing a mathematical and numerical analysis of a given practical problem, it is of capital importance to obtain a unitless formulation of governing equations, due to a number of good reasons. First of all, it enables to determine important scaling parameters (\textsc{Biot} numbers for instance). Henceforth, solving one dimensionless problem is equivalent to solve a whole class of dimensional problems sharing the same scaling parameters. Then, dimensionless equations allow to estimate the relative magnitude of various terms, and thus, eventually to simplify the problem using asymptotic methods \cite{Nayfeh2000}. Finally, the floating point arithmetics is designed such as the rounding errors are minimal if you manipulate the numbers of the same magnitude \cite{Kahan1979}. Moreover, the floating point numbers have the highest density in the interval $(\, 0,\,1 \,)$ and their density decays exponentially when we move further away from zero. So, it is always better to manipulate numerically the quantities at the order of $\O(1)$ to avoid severe round-off errors and to likely improve the conditioning of the problem in hands.

Therefore, the following dimensionless quantities are defined:
\begin{align*}
& u \egal \frac{\Pv}{\Pvi} \,,
&& \uR \egal \frac{\PvR}{\Pvi} \,,
&& \uL \egal \frac{\PvL}{\Pvi} \,,
&& \xs \egal \frac{x}{L} \,, \\[3pt]
& \ts \egal \frac{t}{\tref} \,,
&& \cms \egal \frac{\cm \cdot L^2}{\dmref \cdot \tref} \,,
&& \dms \egal \frac{\dm}{\dmref} \,, 
&& \Pe \egal \frac{\mathsf{v} \cdot L}{\Rv \cdot T \cdot \dmref} \,, \\[3pt] 
& \BivL \egal \frac{\hvL \cdot L}{\dmref}  \,,
&& \BivR \egal \frac{\hvR \cdot L}{\dmref} \,,
&& \glsL \egal \frac{\glL \cdot L}{\dmref \cdot \Pvi} \,.
\end{align*}
In this way, the dimensionless governing equations are then written as:
\begin{subequations}\label{eq:moisture_dimensionlesspb_1D}
\begin{align}
 \cms \pd{u}{\ts} &\egal \pd{}{\xs} \left( \, \dms \ \pd{u}{\xs} \, \right) \moins \Pe \ \pd{u}{\xs} \,,
& \ts & \ > \ 0\,, \;&  \xs & \ \in \ \big[ \, 0, \, 1 \, \big] \,, \\[3pt]
 \dms \, \pd{u}{\xs} &\egal \BivL \cdot \left( \, u \moins \uL \, \right) \moins \glsL \,,
& \ts & \ > \ 0\,, \,&  \xs & \egal 0 \,, \\[3pt]
 -\dms \, \pd{u}{\xs} &\egal \BivR \cdot \left( \, u \moins \uR \, \right)\,,
& \ts & \ > \ 0\,, \,&   \xs & \egal 1 \,, \\[3pt]
 u &\egal 1 \,,
& \ts & \egal 0\,, \,&  \xs & \ \in \ \big[ \, 0, \, 1 \, \big] \,.
\end{align}
\end{subequations}


\section{Numerical schemes}

As the material properties varies with the field, it is not possible to compute an analytical solution of the problem. Therefore, one must use numerical approach, based here on finite differences. It considers a discretisation of the time and space grids with a local approximation of the derivatives. The main issues of a numerical scheme is (i) its global error and (ii) the appropriate behaviour of the solution to represent the physical phenomenon. The primer is quantified by the accuracy of the method, related to the order order of truncation when approximating the derivatives. The second is associated to with the absolute stability of the scheme. A stable scheme avoids to compute a wrong solution. Moreover, an interesting aspect of a numerical scheme is the CPU time, corresponding to the physical clock-time to compute the solution of the problem. Interested readers are invited to consult \cite{Hairer2009, Mendes2016} for more details.

In order to describe numerical schemes, consider a uniform discretisation of the interval $\Ox \ \rightsquigarrow\ \Omega_{\,h}\,$:
\begin{equation*}
  \Omega_{\,h}\ =\ \bigcup_{j\,=\,0}^{N-1} [\,x_{\,j},\;x_{\,j+1}\,]\,, \qquad
  x_{j+1}\ -\ x_{\,j}\ \equiv\ \Delta x\,, \quad \forall j\ \in\ \bigl\{0,\,1,\,\ldots,\,N-1\bigr\}\,.
\end{equation*}
The time layers are uniformly spaced as well $t^{\,n}\ =\ n\,\Delta t\,$, $\Delta t\ =\ \const\ >\ 0\,$, $n\ =\ 0,\,1,\,2,\,\ldots, \, N_{\,t}$. The values of function $u(x,\,t)$ in discrete nodes will be denoted by $u_{\,j}^{\,n}\ \eqdef\ u\,(x_{\,j},\,t^{\,n}\,)\,$.

For the sake of simplicity and without loosing the generality, the numerical schemes are explained for the one-dimensional linear convection equation written as:
\begin{subequations}
\begin{align}\label{eq:conv-diff}
& \pd{u}{t} \plus \pd{J}{x} \egal 0 \,, & t & \ > \ 0\,, \;&  x & \ \in \ \big[ \, 0, \, 1 \, \big]\,, \\
& J \egal a \, u \moins \nu \, \pd{u}{x} \,,
\end{align}
\end{subequations}
where $u(x,\,t)$, $x\ \in\ \Omega\,$, $t\ >\ 0\,$, is the field of interest, $\nu \egal \dfrac{\dm}{\cm} \ >\ 0$ the diffusion coefficient and $a \egal \dfrac{\Pe}{\cm} \in \R$ the advection coefficient. The boundary conditions are also written using a simplified notation:
\begin{align*}
\pd{u}{x} & \egal \mathrm{Bi} \cdot \left( \, u \moins \uL \, \right) \moins \mathrm{g} \,, && x \egal 0 \,, \\
- \ \pd{u}{x} & \egal \mathrm{Bi} \cdot \left( \, u \moins \uR \, \right)\,, && x \egal 1 \,.
\end{align*}
In addition, for the sake of clarity, the upper-script $\star$ standing for dimensionless parameter is no longer used.

\begin{figure}
  \centering
  \def\svgwidth{0.70\textwidth}
  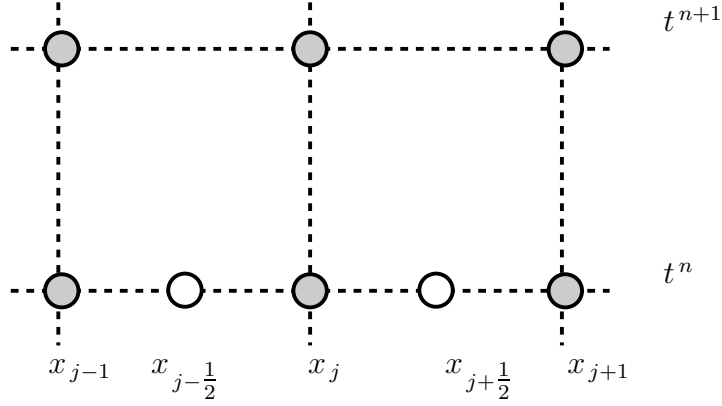
  \caption{\small\em Stencil of the numerical schemes.}
  \label{fig:stencil}
\end{figure}


\subsection{The \CN ~scheme}

The method proposed by \textsc{Crank} \& \textsc{Nicolson} (CN) \cite{Crank1947} is widely popular in many applications, and it is known specially for its stability. In this case, the CN scheme is applied to the convection--diffusion equation~\eqref{eq:conv-diff}:
\begin{align}\label{eq:cn}
&  \frac{u_{\,j}^{\,n+1}\ -\ u_{\,j}^{\,n}}{\Delta t}\ +\ \frac{1}{\Delta x}\;\biggl[\, J_{\,j+\frac{1}{2}}^{\,n+\frac{1}{2}}\ -\ J_{\,j-\frac{1}{2}}^{\,n+\frac{1}{2}} \, \biggr]\ =\ 0\,, & j\ =\ 1,\,\ldots,\,N-1\,, \qquad n\ \geqslant\ 0\,,
\end{align}
in which,
\begin{align}\label{eq:flux_CN}
J_{\,j \pm \frac{1}{2}}^{\,n + \frac{1}{2}} \ =\ \dfrac{1}{2} \ \Bigl(\, J_{\,j \pm \frac{1}{2}}^{\,n} \ +\ J_{\,j \pm \frac{1}{2}}^{\,n+1} \,\Bigr) \,.
\end{align}
The flux $J$ is defined using the upwind scheme \cite{Patankar1980}:
\begin{align}\label{eq:up-wind}
J_{\,j + \frac{1}{2}}^{\,n}\ =\ \left. \begin{cases} 
a \ u_{\,j}^{\,n}\,, & a \geqslant 0 \\
a \ u_{\,j+1}^{\,n}\,, & a < 0 
\end{cases}
\ \right\} -\ \nu\,\dfrac{\bigl(\, u_{\,j+1}^{\,n}\ -\ u_{\,j}^{\,n} \,\bigr)}{\Delta x} \,. 
\end{align}
Parameter $a$ is related to the advection transfer mechanisms in Eq.~\eqref{eq:conv-diff}. Thus, an upwind scheme defines the flux $J$ according to the sense of the advection transfer. If $a \ \geqslant \ 0$, the advective flux is directed from $x_{\,j}$ to $x_{\,j+1}$ and the flux is therefore approximated using $u_{\,j}^{\,n}$.

Substituting Eq.~\eqref{eq:up-wind} into Eq.~\eqref{eq:cn}, we obtain a discrete dynamical system:
\begin{multline}
 \bigl[\, 1 \plus 2\,\lambda \plus \gamma \ (\,\bp \moins \bm \,)\,\bigr]\, u_{\,j}^{\,n+1} 
 \moins (\, \lambda \moins \gamma\,\bm\, )\, u_{\,j+1}^{\,n+1} 
 \moins (\, \gamma\,\bp \plus \lambda\, )\, u_{\,j-1}^{\,n+1} \egal \\[3pt]
 \bigl[\, 1 \moins 2\,\lambda \moins \gamma \ (\,\bp \moins \bm \,)\,\bigr]\, u_{\,j}^{\,n} 
 \plus (\, \lambda \moins \gamma\,\bm\, )\, u_{\,j+1}^{\,n} 
 \plus (\, \gamma\,\bp \plus \lambda\, )\, u_{\,j-1}^{\,n} \,,
\end{multline}
where
\begin{align*}
& \lambda\ =\ \dfrac{\nu\,\Delta t}{2\,\Delta x^{\,2}}\,, & & \gamma\ =\ \dfrac{a\,\Delta t}{2\,\Delta x} & & \text{and}  & &
\bpm\ =\ \dfrac{1\ \pm\ \sign(a)}{2} \,.
\end{align*}
This scheme is $\O(\Delta t^2\ +\ \Delta x)$ accurate and unconditionally stable (at least for linear problems). The loose of one order in $\dx$ is due to the advection. If $\Pe \egal 0$, than $\O(\Delta t^2\ +\ \Delta x^2)$ . Its stencil is illustrated in Figure~\ref{fig:stencil}.


\subsection{The \SG ~scheme}

The straightforward discretisation of Eq.~\eqref{eq:conv-diff} by  central differences yields to the following equation:
\begin{align*}
\frac{u_{\,j}^{\,n+1}\ -\ u_{\,j}^{\,n}}{\Delta t} 
\plus \frac{1}{\Delta x} \Biggl[\, J_{\,j+\frac{1}{2}}^{\,n}\ -\ J_{\,j-\frac{1}{2}}^{\,n}   \, \Biggr] 
\egal 0 \,.
\end{align*}
Assuming $J$ constant on the dual cell $\bigl[ \, x_{j}, \, x_{j+1}\, \bigr]$, \textsc{Scharfetter} and \textsc{Gummel} start from the fact that the numerical flux at each interface $x_{\,j+\half}$ can be computed giving the following boundary-value problem \cite{Scharfetter1969, Gosse2013, Gosse2017}:
\begin{subequations}\label{sec1_eq:SP_equation}
\begin{align}
 J_{\,j+\half}^{\,n} & \egal a \, u \moins \nu \, \pd{u}{x} \,, & & \forall x \ \in \ \big[ \, x_{j}, \, x_{j+1} \, \big] \,; \\
 u & \egal u_{\,j}^{\,n} \,, && x \egal x_{j} \,;\\
 u  &\egal u_{\,j+1}^{\,n} \,, && x \egal x_{j+1} \,.
\end{align}
\end{subequations}
An advantageous feature is that solution of Eq.~\eqref{sec1_eq:SP_equation} corresponds to the solution of the \textsc{Poincar\'e}--\textsc{Steklov} operator,
\begin{align*}
\mathcal{S}\,:\ (u_{\,j}^{\,n},\,u_{\,j+1}^{\,n})\ \mapsto\ J_{\,j+\half}^{\,n}
\end{align*}
and can be expressed as:
\begin{align}\label{sec1_eq:SP_solution}
& J_{\,j+\half}^{\,n}  \egal \frac{\nu}{\Delta x} \Biggl[ \, -\ \Bp \, u_{\,j+1}^{\,n} \plus \Bm u_{\,j}^{\,n} \, \Biggr] \,,
\end{align}
where $\mathcal{B}$ corresponds to the \textsc{Bernoulli} function defined by
\begin{align*}
  \mathcal{B}(\,z \,) \ \eqdef \ \frac{z}{\mathrm{e}^{\,z} \moins 1} \,.
\end{align*}
Therefore, given Eq.~\eqref{sec1_eq:SP_solution} and the \textsc{Poincar\'e}--\textsc{Steklov} operator, the \SG ~numerical ~scheme is written as:
\begin{align*}
&  \frac{u_{\,j}^{\,n+1}\ -\ u_{\,j}^{\,n}}{\Delta t}\ +\ \frac{1}{\Delta x}\;  \Bigl[\, \mathcal{S}\bigl(u_{j+1}^{n}, u_j^{n} \bigr) - \mathcal{S} \bigl( u_{j}^{n}, u_{j-1}^{n} \bigr) \, \Bigr]\ =\ 0\,.
\end{align*}
Finally, the scheme yields to:
\begin{align*}
u_{\,j}^{\,n+1} & \egal u_{\,j}^{\,n}
\plus \lambda \, \Biggl[ \, \mathcal{B} \bigl(\, \theta \,\bigr) \, u_{\,j+1}^{\,n}
\moins \biggl(\, \mathcal{B} \bigl(\, - \, \theta \,\bigr) 
\plus \mathcal{B} \bigl(\, \theta \,\bigr) \,\biggr) \, u_{\,j}^{\,n} 
\plus \mathcal{B} \bigl(\, - \, \theta \,\bigr) \, u_{\,j-1}^{\,n} 
\, \Biggr]\,, 
\end{align*}
where 
\begin{align*}
& \lambda \ \eqdef \ \dfrac{ \nu \ \Delta t}{2 \ \Delta x^{\,2}} \, && \text{and} && \ \theta \ \eqdef \ \dfrac{a \ \Delta x}{\nu} \,.
\end{align*}
Another interesting point is that, considering Eq.~\eqref{sec1_eq:SP_equation}, the exact interpolation of solution $u(x)$ can be computed for $x \ \in \ \big[ \, x_{j}, \, x_{j+1} \, \big]$ by:
\begin{align*}
& u(x) \egal \frac{1}{a} \ J_{\,j+\half}^{\,n} \plus 
\frac{\bigr( \, u_{\,j}^{\,n} \moins u_{\,j+1}^{\,n} \, \bigr)}{\mathrm{e}^{\,\dfrac{a \ x_{\,j}}{\nu}} \moins  \mathrm{e}^{\,\dfrac{a \ x_{\,j+1}}{\nu}}}\,
\mathrm{e}^{\,\dfrac{a \, x}{\nu}} \,.
\end{align*}

For the nodes at the boundary surface, $j \egal \big\{\, 1, \, N \, \big\} \,$, the flux $J_{\,\half}$ is solution of
\begin{align*}
 J_{\,\half}^{\,n} & \egal a \, u \moins \nu \, \pd{u}{x} \,, & & \forall x \in \big[ \, 0, \, x_{\,1} \, \big] \,; \\
\pd{u}{x} &\egal \mathrm{Bi} \cdot \left( \, u \moins \uL \, \right) \moins \mathrm{g} \,, && x \egal 0 \,;\\
 u  &\egal u_{\,1}^{\,n} \,, && x \egal x_{1} \,.
\end{align*}
and $J_{\,N+\half}$ of
\begin{align*}
 J_{\,N+\half}^{\,n} & \egal a \, u \moins \nu \, \pd{u}{x} \,, & & \forall x \in \big[ \, x_{\,N}, \, 1 \, \big] \,; \\
  u  & \egal u_{\,N}^{\,n} \,, && x \egal x_{N} \,; \\
  - \ \pd{u}{x} &\egal \mathrm{Bi} \cdot \left( \, u \moins \uR \, \right) \,, && x \egal 1 \,.
\end{align*}
Thus, we have:
\begin{align*}
  J_{\,\half}^{\,n} & \egal a \ \frac{\bigl( \, \mathrm{Bi} \moins a \, \bigr) \,u_{\,1}^{\,n} \moins \bigl(\, \mathrm{Bi} \, \uL \plus g \,\bigr) \, \mathrm{e}^{\,\dfrac{a \ \dx}{2\, \nu}}}{\mathrm{Bi}  \moins a \moins \mathrm{Bi} \, \mathrm{e}^{\,\dfrac{a \ \dx}{2\, \nu}}} \,, \\
  J_{\,N+\half}^{\,n} & \egal a \ \frac{\mathrm{Bi} \, \uR \moins \bigl(\, \mathrm{Bi}  \moins a \,\bigr) u_{\,N}^{\,n} \, \mathrm{e}^{\,\dfrac{a \ \dx}{2\, \nu}}}{\mathrm{Bi} \moins \bigl(\, \mathrm{Bi}  \moins a \,\bigr) \, \mathrm{e}^{\,\dfrac{a \ \dx}{2\, \nu}}} \,.
\end{align*}

The stencil is illustrated in Figure~\ref{fig:stencil}. The important feature of the \SG ~numerical scheme is well balanced as well as asymptotically preserved. The limiting behaviour of the discrete equations is correct independently from grid parameters:
\begin{align*}
& \lim_{a \ \rightarrow \ 0} J_{\,j+\half}^{\,n} \egal - \ \nu\;\frac{u_{\,j+1}^{\,n} \moins u_{\,j}^{\,n}}{\Delta x} \,, \\
& \lim_{\nu \ \rightarrow \ 0} J_{\,j+\half}^{\,n} \egal  
\begin{cases} 
a\,u_{\,j}^{\,n} \,, & \quad  a \ \leqslant \ 0 \,, \\
a\,u_{\,j+1}^{\,n} \,, & \quad  a \ > \ 0 \,.
\end{cases}
\end{align*}
Furthermore, the computation of $J_{\,j+\half}^{\,n}$ is exact and it gives an excellent approximation of the physical phenomena. The only approximation is done when assuming $J_{\,j+\half}^n$ constant in the interval $\Bigl[ \, x_{\,j-\half}^{\,n}, \, x_{\,j+\half}^{\,n}\, \Bigr]$. In addition, when the steady state is reached, the solution computed with the scheme becomes exact \cite{Jerome1991}. Interested readers may consult \cite{Patankar1980, Gosse2013, Gosse2017} for recent works on the \SG ~scheme.

On the contrary to the \CN, the \SG ~scheme is not unconditionally stable. It has a stability limitation. It means that the scheme can compute a solution when the so called standard \textsc{Courant--Friedrichs--Lewy} (CFL) condition is respected. In the linear case, the CFL condition of the \SG ~scheme is given by \cite{Gosse2016}: 
\begin{align}\label{sec1_eq:cfl_SG}
\dt \, \dfrac{\Pe}{\dms} \, \tanh \left(\, \dfrac{\Pe \, \dx}{2 \, \dms} \,\right)^{-1} \ \leqslant\ \dx \, \frac{\cms}{\dms} \,.
\end{align}
If the spatial grid is refined, $\lim\limits_{\dx \rightarrow 0} \tanh (\dx)  \egal \dx\,$, and the CFL condition starts to become quadratic $\dt \leq C_1 \cdot \dx^2\,$. However, if the spatial grid is large, $\lim\limits_{\dx \rightarrow 1} \tanh (\dx) \egal 1$ and the CFL condition yields to $\dt \leq C_2 \cdot \dx$. For these reasons, the values of $\dx$ have to be in a closed interval, depending on the material properties.


\subsection{Comparison of numerical schemes}

A primary comparison of the numerical schemes can be done by computing the $\mathcal{L}_{\,2}$ error between the solution $u_{\, \mathrm{num}}$ and a reference solution $u_{\, \mathrm{ref}}\,$:
\begin{align}
\varepsilon & \eqdef\ \sqrt{\,\frac{1}{N_{\,t} \ N_{\,x}} \, \sum_{n=1}^{N_t} \  \sum_{j=1}^{N_x}\, \Bigl(\, \, u_{\mathrm{num}}(x_{\,j},t_{\,n}) \moins u_{\mathrm{ref}}(x_{\,j},t_{\,n}) \, \Bigr)^2}\,.
\end{align}
The reference solution is computed using the \texttt{Matlab} open source package \emph{Chebfun} \cite{Driscoll2014}. Using the function \texttt{pde23t}, it enables to compute a numerical solution of a partial derivative equation using the \textsc{Chebyshev} functions. The $\mathcal{L}_{\,2}$ error can also be computed along the space or time domains, according to: 
\begin{align*}
& \varepsilon(\,x\,) \ \eqdef\ \sqrt{\,\frac{1}{N_{\,t}} \, \sum_{n=1}^{N_t} \, \Bigl(\, \, u_{\mathrm{num}}(x,t_{\,n}) \moins u_{\mathrm{ref}}(x,t_{\,n}) \, \Bigr)^2} \,, \\
& \varepsilon(\,t\,) \ \eqdef\ \sqrt{\,\frac{1}{N_{\,x}} \ \sum_{j=1}^{N_x}\, \Bigl(\, \, u_{\mathrm{num}}(x_{\,j},t) \moins u_{\mathrm{ref}}(x_{\,j},t) \, \Bigr)^2}\,.
\end{align*}


\section{Numerical application: linear case}

A first case of linear moisture transfer is considered to validate the solutions obtained by the \CN ~and \SG ~numerical schemes. The dimensionless properties of the material are equal to $\dms \egal 1$ and $\cms \egal 47 \,$. The \textsc{P\'eclet} Number is taken as $\PE \egal 20 \,$. The final simulation time is fixed to $\tau^{\,\star} \egal 120 \,$. The external and internal \textsc{Biot} numbers are $\BivL \egal 2.5$ and $\BivR \egal 1$, respectively. The boundary conditions were described as:
\begin{align*}
& \uL (\ts) \egal 1 \plus \frac{1}{2} \; \sin \left(\, \frac{2\pi \, \ts}{24}\,\right) 
\plus \frac{3}{10} \; \sin \left(\, \frac{2\pi \, \ts}{4}\,\right) \,,\\
& \uR (\ts) \egal 1 \plus \frac{4}{5} \;\sin \left(\, \frac{2\pi \, \ts}{12}\,\right) \,.
\end{align*}

\begin{figure}
\centering
\includegraphics[width=0.65\textwidth]{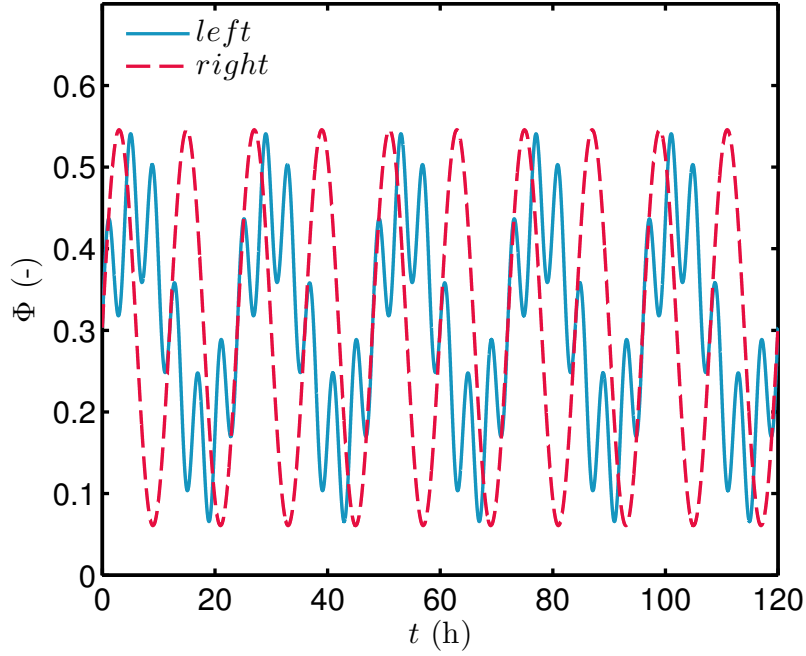}
\caption{\small\em Boundary conditions.}
\label{fig_AN1:BC}
\end{figure}

From a physical point of view, the numerical values correspond to a material length $L \egal 0.1$~$\mathsf{m}$. The moisture properties are $\dm \egal 3 \dix{-10} $ $\mathsf{s}$ and $\cm \egal 1.8 \dix{-4}$ $\mathsf{kg/m^3/s}$, corresponding approximately to the wood fibre from \cite{Rafidiarison2015}. The initial vapour pressure in the material is considered uniform $\Pvi \egal 7.1 \dix{2}$ $\mathsf{Pa \,}$, corresponding to a relative humidity of $30$ $\mathsf{\%}$. The reference time is $\tref \egal 1$  $\mathsf{h}$, thus the total time of simulation corresponds to $120$ hours, or five days. The boundary conditions, represented by the relative humidity $\phi$ are given in Figure~\ref{fig_AN1:BC}. The sinusoidal variations oscillate between dry and moist state during 120 hours. The convective vapour coefficients are set to $8 \dix{-9}$  $\mathsf{s/m}$ and $3.5 \dix{-9}$  $\mathsf{s/m}$ for the left and right boundary conditions, respectively.

\begin{figure}
\centering
\subfigure[a][\label{fig_AN1:x0time}]{\includegraphics[scale=.6]{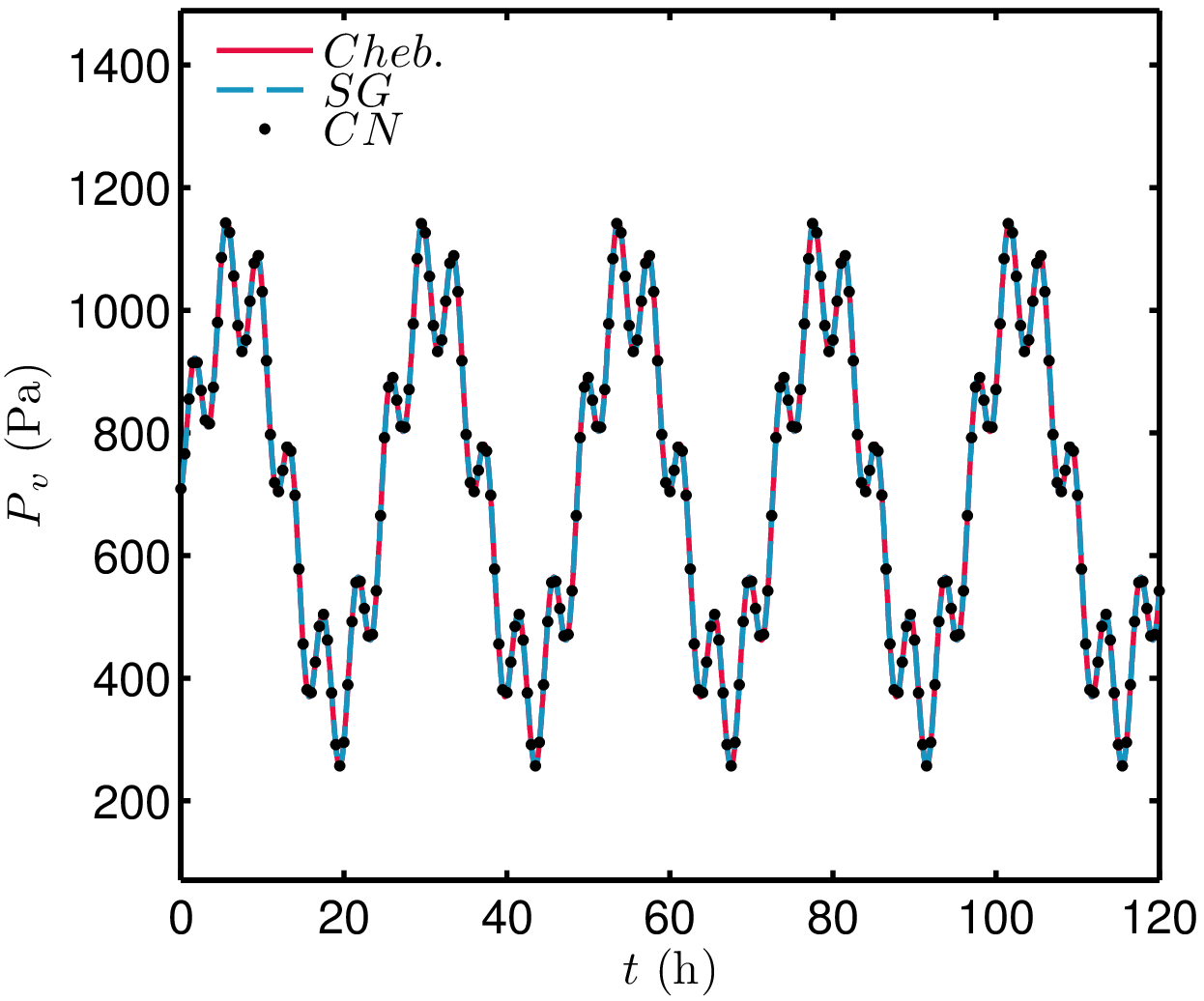}} \hspace{0.3cm}
\subfigure[b][\label{fig_AN1:profil}]{\includegraphics[scale=.6]{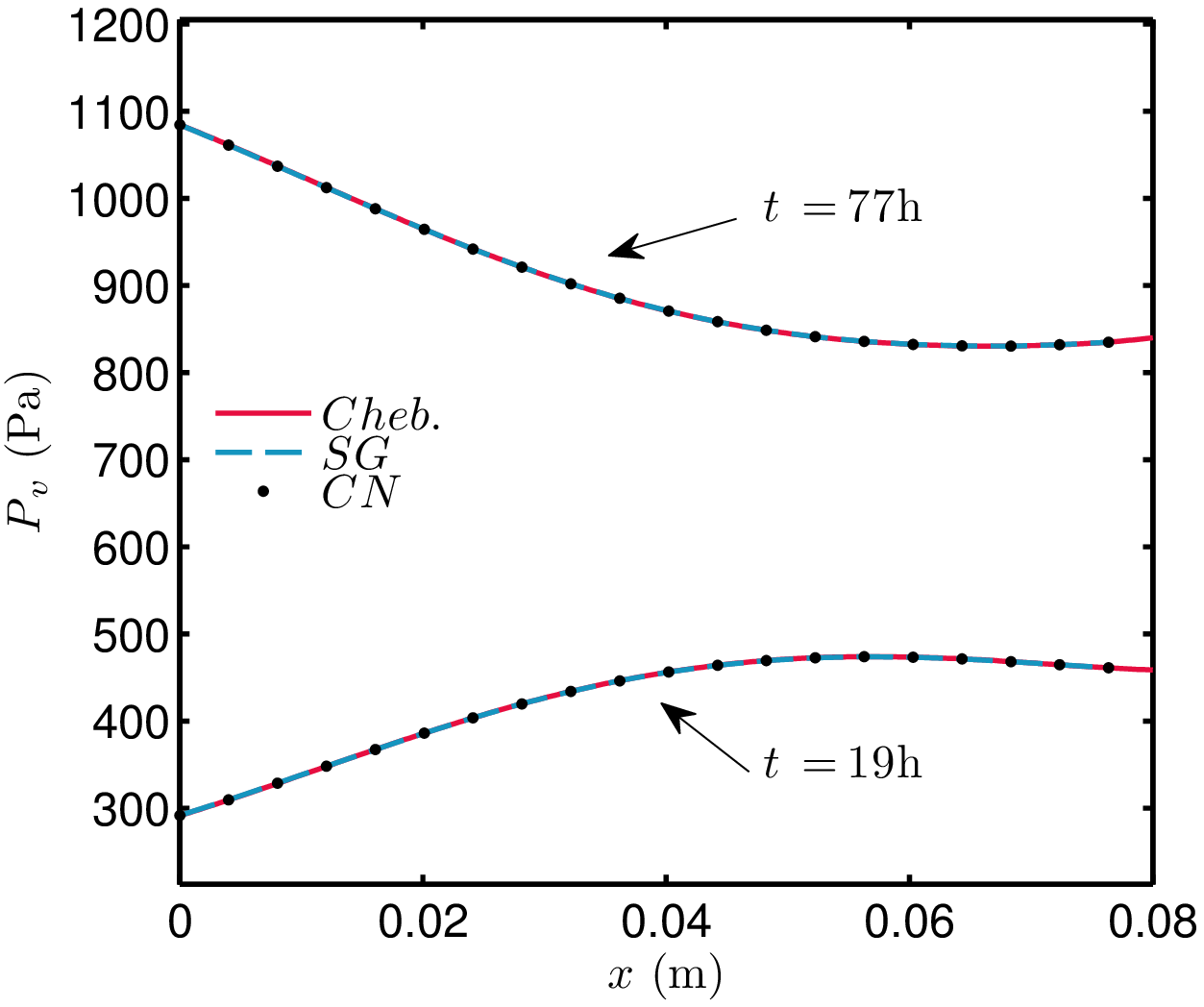}}
\caption{\small\em Vapour pressure time evolution at $x \egal 0$  $\mathsf{m}$ (a) and profiles for $t \in \left\lbrace 19, \, 77\right\rbrace$  $\mathsf{h}$ (b).}
\end{figure}

The solution of the problem has been first computed for a discretisation $\Delta \xs \egal 5 \cdot 10^{-4}$ and $\Delta \ts \egal 10^{-3} \,$. The physical phenomena are thus well represented, as illustrated in Figure~\ref{fig_AN1:x0time} with the time evolution of the vapour pressure at $x \egal 0 \,$. The variations follow the ones of the left boundary conditions. It can be noted a good agreement between the two numerical schemes with the reference. Furthermore, the vapour pressure profile is shown in Figure~\ref{fig_AN1:profil} for $t \egal 19$~$\mathsf{h\,}$ and  $t \egal 77$~$\mathsf{h\,}$. 

\begin{figure}
\centering
\subfigure[a][\label{fig_AN1:err_ft}]{\includegraphics[scale=.6]{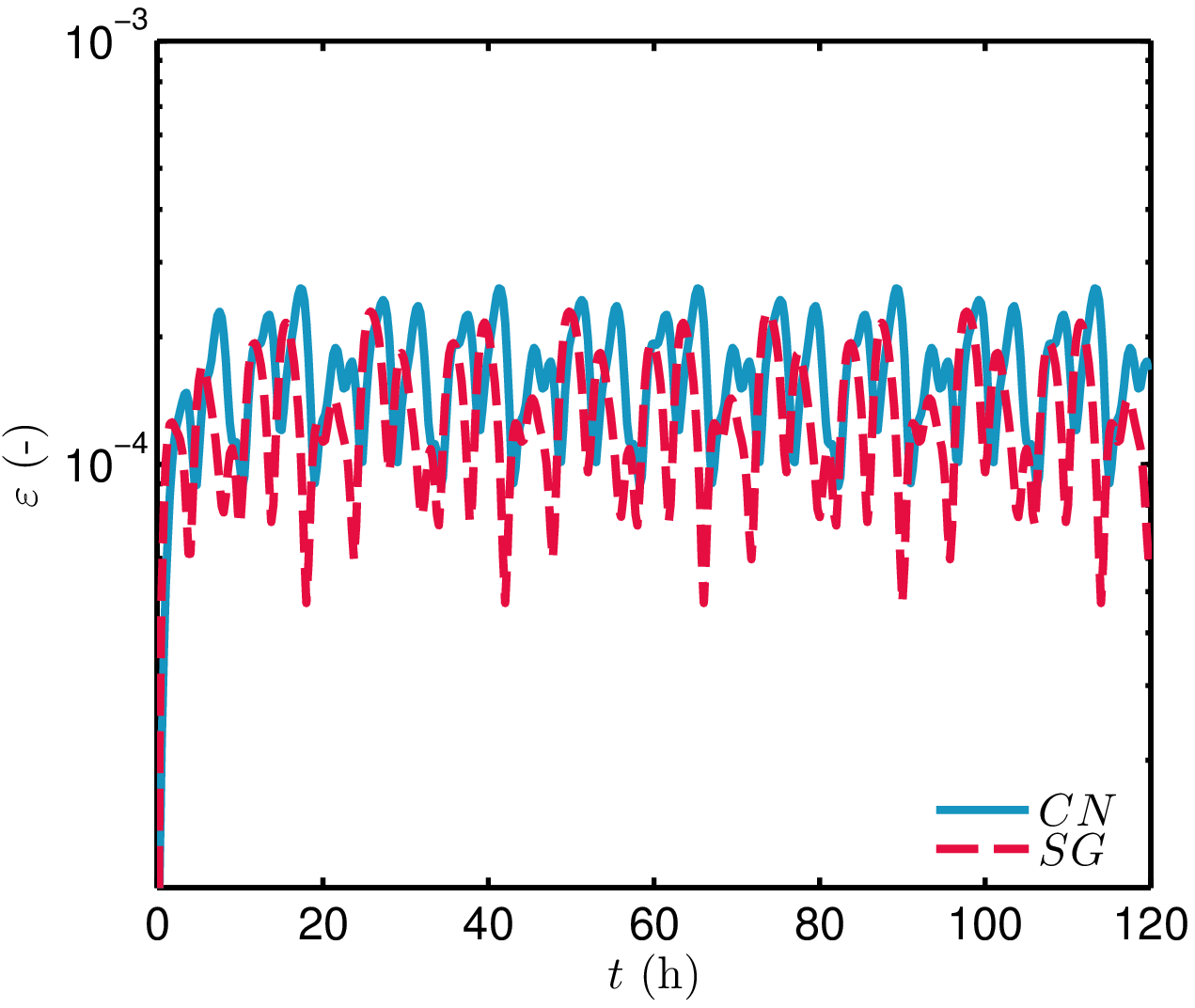}} \hspace{0.3cm}
\subfigure[b][\label{fig_AN1:err_fx}]{\includegraphics[scale=.6]{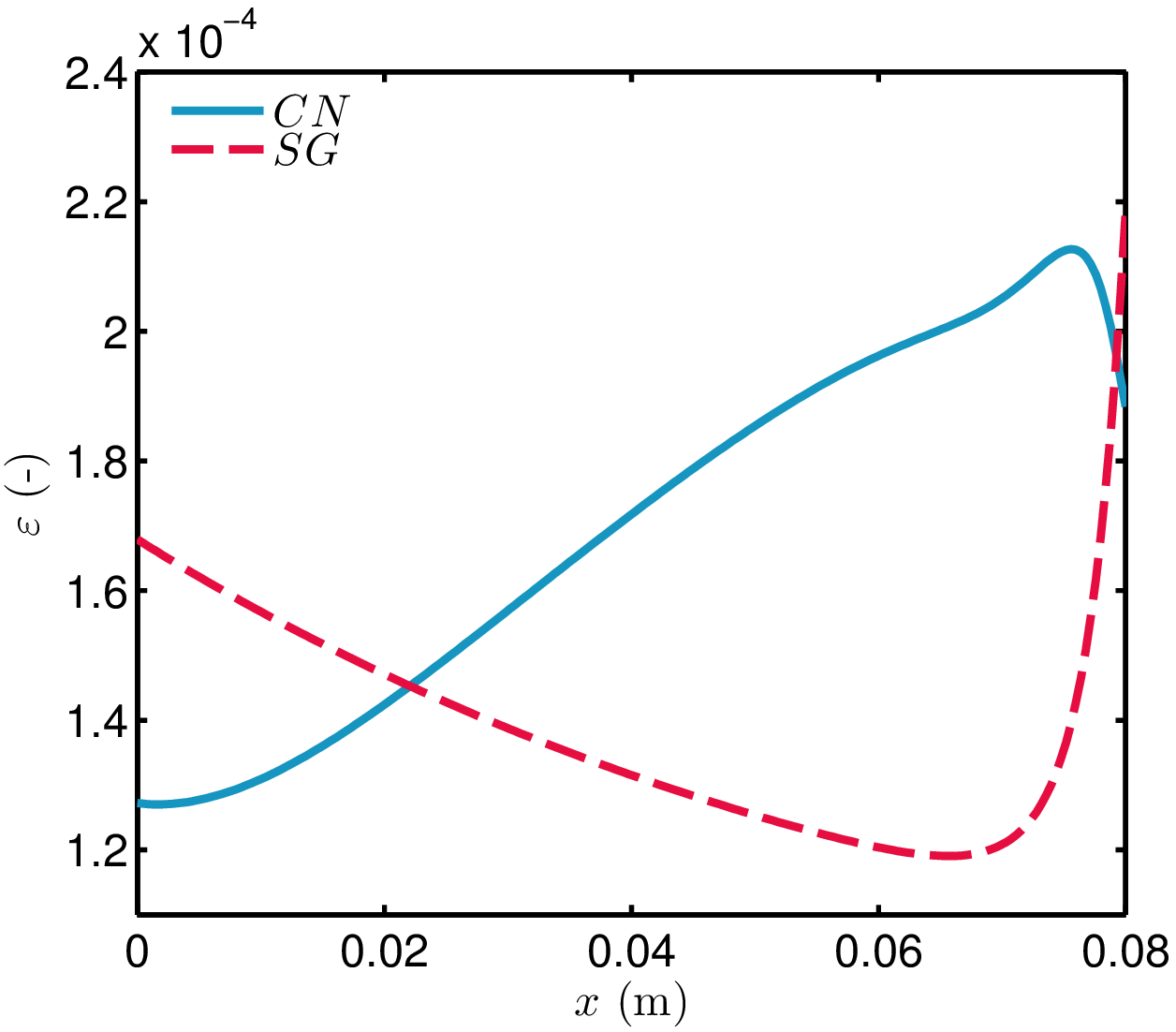}}
\caption{\small\em $L_{\,2}$ error for fixed $\Delta \xs \egal 10^{\,-4}$ and $\Delta \ts \egal 10^{\,-3}$, in function of $t$ (a) and in function of $x$ (b).}
\end{figure}

The reference solution has been computed using the \texttt{Matlab} open source toolbox \emph{Chebfun} \cite{Driscoll2014}. Using the function \texttt{pde23t}, it enables to compute a numerical solution of a partial derivative equation using the \textsc{Chebyshev} functions. Both \CN ~and \SG ~numerical schemes give accurate results as illustrated with the $\mathcal{L}_{\,2}$ error calculated as a function of $x$ in Figure~\ref{fig_AN1:err_fx} and calculated as a function of $t$ in Figure~\ref{fig_AN1:err_ft}. In these figures it is possible to verify that the order of accuracy of both schemas is $\O(\Delta t^2\ +\ \Delta x)$. In this case study, it was considered $\Delta x^{\,\star} \egal 5 \cdot 10^{-4}$ and $\Delta t^{\,\star} \egal 10^{-3}$, which leads to an order of accuracy of $\O(10^{\,-6}\ +\ 5 \cdot 10^{-4}) \sim \O(10^{-4})$.

\begin{figure}
\centering
\subfigure[a][\label{fig_AN1:err_fdt}]{\includegraphics[scale=.6]{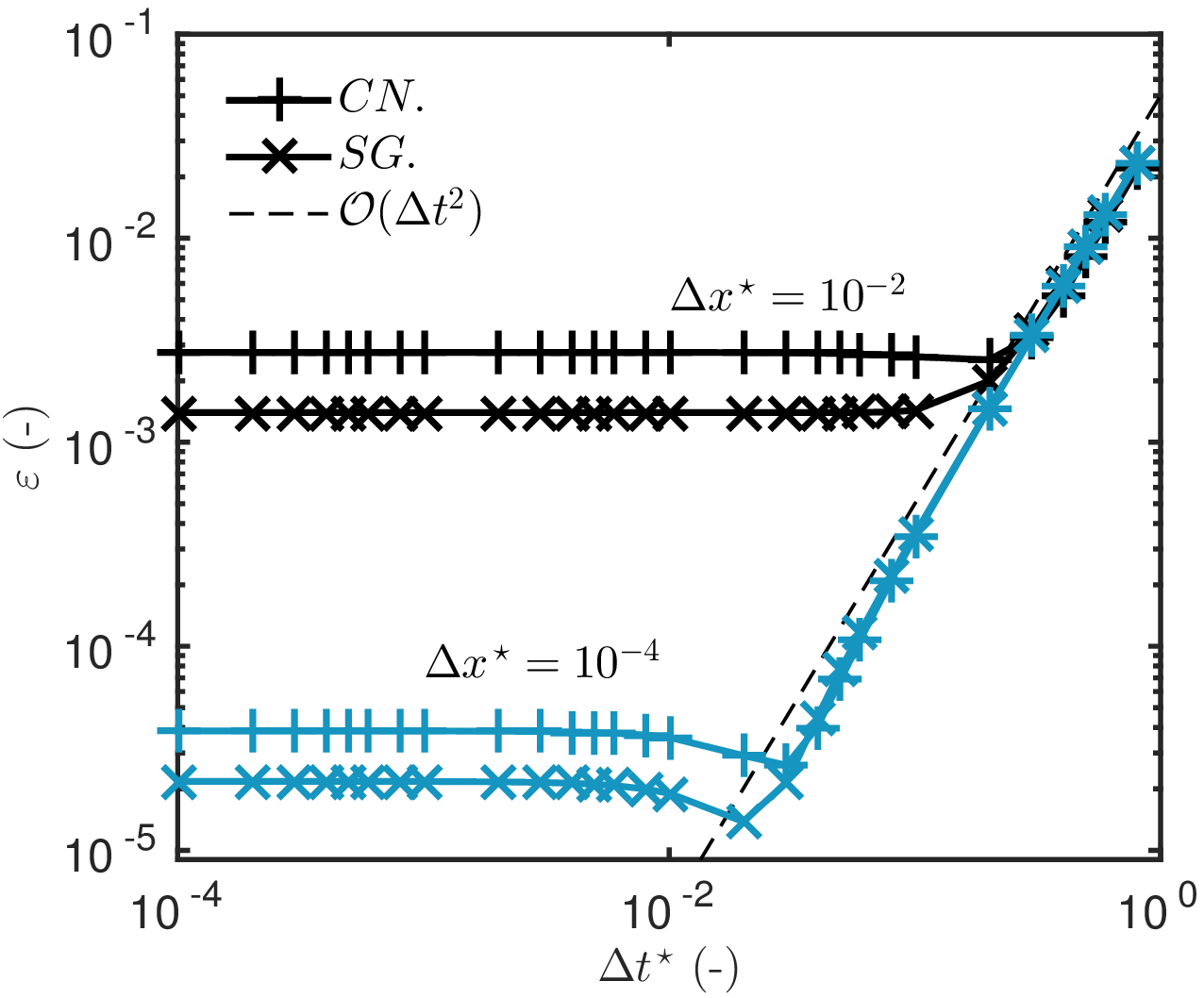}} \hspace{0.3cm}
\subfigure[b][\label{fig_AN1:err_fdx}]{\includegraphics[scale=.6]{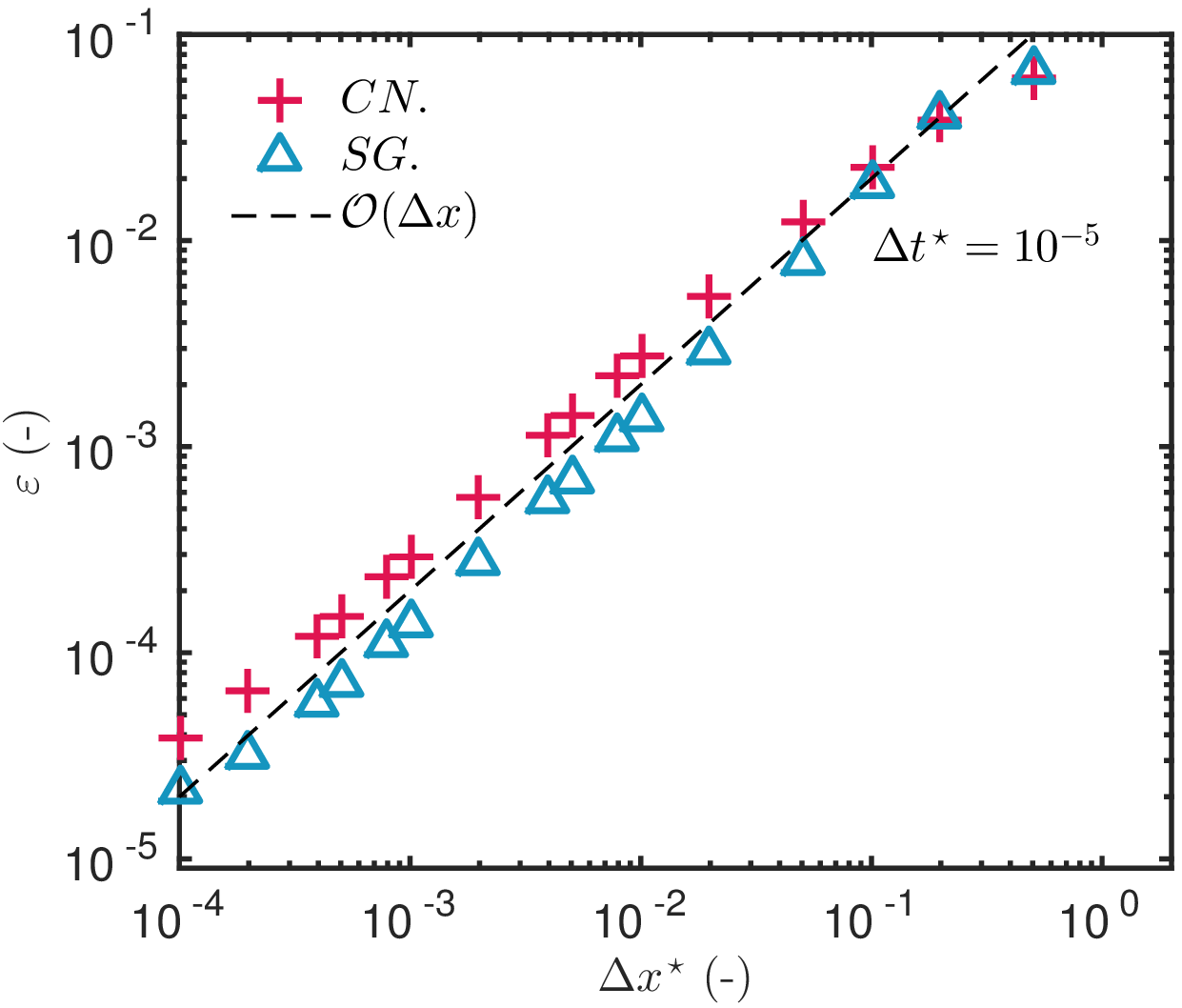}}
\caption{\small\em $\mathcal{L}_{\,2}$ error as a function of $\Delta \ts$ for the CN and SG schemes ($\Delta \xs \egal 10^{-4}$) (a) and $\mathcal{L}_{\,2}$ error as a function of $\Delta \xs$ for the CN and SG schemes ($\Delta \ts \egal 10^{-5}$).}
\end{figure}

A numerical analysis of the behaviour of the two numerical schemes has been carried out for different values of the temporal discretisation $\Delta \ts$ and spatial discretisation $\Delta \xs$. The spatial discretisation is maintained to $\Delta \xs \egal 10^{\,-4}$ and $\Delta \xs \egal 10^{\,-2} \,$. Results of $\mathcal{L}_{\,2}$ error can be seen in Figure~\ref{fig_AN1:err_fdt} and Figure~\ref{fig_AN1:err_fdx}. It can be seen in Figure~\ref{fig_AN1:err_fdt} that the error has a minimum value when $\Delta \ts$ varies because the order of error of  $\Delta \xs$ is higher.  After that minimum value, the error gets proportional to a constant multiplied by the order of $\O(\,\Delta t^2\,)$. Both figures confirm that the errors of the \SG ~and \CN ~schemes are proportional to $\O(\,\Delta t^2\,)$ and $\O(\,\Delta x\,)$, and that the SG scheme has some advantages compared to the CN scheme.

\section{Extension for nonlinear moisture transfer}

The previous case study investigated the use of the numerical schemes for computing the solution of a linear problem of moisture convection. This second case study considers now nonlinear transfer, due to diffusion material properties depending on the moisture content $\dms\left(\, u \, \right)$ and  $\cms\left(\, u \, \right)$. The \textsc{Peclet} number $\Pe$ ~is considered as constant, assuming a constant mass average velocity and temperature inside the material. In the next section, it will be considered as variable. This case will be investigated by means of the \SG ~and an improved version of the \CN ~schemes. First, the \CN ~and \SG ~schemes are detailed for the nonlinear case. For this, Eq.~\eqref{eq:moisture_dimensionlesspb_1D} is re-called with a simplified notation:
\begin{subequations}
\begin{align}\label{eq:conv-diff_NL}
& c\,(u) \, \pd{u}{t} \plus \pd{J}{x} \egal 0 \,, & t & \ > \ 0\,, \;&  x & \ \in \ \big[ \, 0, \, 1 \, \big] \,,\\
& J \egal \Pe\,(u) \, u \moins d\,(u) \, \pd{u}{x} \,.
\end{align}
\end{subequations}


\subsection{The \CN ~scheme and its improved version}

The straightforward application of the \CN ~scheme to Eq.~\eqref{eq:conv-diff_NL} yields the following scheme:
\begin{align}\label{eq:CN_NL}
c_{\,j}^{\,n} \, \frac{u_{\,j}^{\,n+1}\ -\ u_{\,j}^{\,n}}{\Delta t} 
\egal \frac{1}{\Delta x} \biggl[\, J_{\,j+\frac{1}{2}}^{\,n+\frac{1}{2}}\ -\ J_{\,j-\frac{1}{2}}^{\,n+\frac{1}{2}}   \, \biggr] \,.
\end{align}
Considering Eq.~\eqref{eq:flux_CN}, we get: 
\begin{align*}
J_{\,j + \frac{1}{2}}^{\,n+1}\ & =\ \left. \begin{cases}
\Pe_{\,j}^{\,n+1} \ u_{\,j}^{\,n+1}\,, & \Pe_{\,j}^{\,n+1} \geqslant 0 \\
\Pe_{\,j+1}^{\,n+1} \ u_{\,j+1}^{\,n+1}\,, & \Pe_{\,j+1}^{\,n+1} < 0 
\end{cases}
\ \right\} -\ \frac{1}{2} \ \left(\, d \; \pd{u}{x} \, \right)_{\,j+\half}^{\,n+1}  \\
& =\ \left. \begin{cases}
\Pe_{\,j}^{\,n+1} \ u_{\,j}^{\,n+1}\,, & \Pe_{\,j}^{\,n+1} \geqslant 0 \\
\Pe_{\,j+1}^{\,n+1} \ u_{\,j+1}^{\,n+1}\,, & \Pe_{\,j+1}^{\,n+1} < 0 
\end{cases}
\ \right\} -\ \frac{1}{2 \ \Delta x} \ d_{\,j+\half}^{\,n+1} \, \Big(\, u_{\,j+1}^{\,n+1} \moins u_{\,j}^{\,n+1} \,\Big)  \,.
\end{align*}
The numerical scheme is then written as:
\begin{multline*}
 \biggl[\, 1 \plus \gamma_{\,j+\half}^{\,n+1} \ (\,\bp \moins \bm \,) \plus \lambda_{\,j+\half}^{\,n+1} \plus \lambda_{\,j-\half}^{\,n+1} \,\biggr]\, u_{\,j}^{\,n+1} \\ 
 \moins \biggl[\, \lambda_{\,j+\half}^{\,n+1} \moins \gamma_{\,j+\half}^{\,n+1} \,\bm\, \biggr]\, u_{\,j+1}^{\,n+1} 
 \moins \biggl[\, \gamma_{\,j-\half}^{\,n+1} \,\bp \plus \lambda_{\,j-\half}^{\,n+1} \, \biggr]\, u_{\,j-1}^{\,n+1} \egal \\ \biggl[\, 1 \moins \gamma_{\,j+\half}^{\,n} \ (\,\bp \moins \bm \,) \moins \lambda_{\,j+\half}^{\,n} \plus \lambda_{\,j-\half}^{\,n} \,\biggr]\, u_{\,j}^{\,n} 
 \plus \\ \biggl[\, \lambda_{\,j+\half}^{\,n} \moins \gamma_{\,j+\half}^{\,n} \,\bm\, \biggr]\, u_{\,j+1}^{\,n} 
 \plus \biggl[\, \gamma_{\,j-\half}^{\,n} \,\bp \plus \lambda_{\,j-\half}^{\,n} \, \biggr]\, u_{\,j-1}^{\,n} \,,
\end{multline*}
where
\begin{align*}
& \lambda_{\,j}\ =\ \dfrac{d_{\,j}\,\Delta t}{2\,\Delta x^{\,2} \, c_{\,j}}\,, & & \gamma_{\,j}\ =\ \dfrac{ \Pe_{\,j}\,\Delta t}{2\,\Delta x} & & \text{and}&&
\bpm\ =\ \dfrac{1\pm\sign(\Pe)}{2} \,.
\end{align*}
However, this approach leads to deal with nonlinearities associated to the quantities of the flux (as $d_{\,j+\half}^{\,n+1}$ and $\Pe_{\,j}^{\,n+1}$) at the upcoming time layer $t \egal t^{\,n+1}$. To deal with this issue, linearisation techniques as \textsc{Picard} or \textsc{Newton}--\textsc{Raphson} ones \cite{Raphson1690, Cajori1911}, can be employed but requiring a high number of sub-iterations.

To overcome these difficulties, it is possible to evaluate the diffusion coefficient at the current time layer instead of the upcoming one \cite{Ascher1995}. Thus, the convection flux at the interface becomes:
\begin{align*}
J_{\,j + \frac{1}{2}}^{\,n+1}\ & =\ \left. \begin{cases}
\Pe_{\,j}^{\,n} \ u_{\,j}^{\,n+1}\,, & \Pe_{\,j}^{\,n} \geqslant 0 \\
\Pe_{\,j}^{\,n} \ u_{\,j+1}^{\,n+1}\,, & \Pe_{\,j}^{\,n} < 0 
\end{cases}
\ \right\} -\ \frac{1}{2 \ \Delta x} \ d_{\,j+\half}^{\,n} \, \Big(\, u_{\,j+1}^{\,n+1} \moins u_{\,j}^{\,n+1} \,\Big)  \,.
\end{align*}
Finally, the \mCN ~schemes yields to: 
\begin{multline}\label{eq:schema_CN_NL}
 \biggl[\, 1 \plus \gamma_{\,j+\half}^{\,n} \ (\,\bp \moins \bm \,) \plus \lambda_{\,j+\half}^{\,n} \plus \lambda_{\,j-\half}^{\,n} \,\biggr]\, u_{\,j}^{\,n+1} \\
 \moins \biggl[\, \lambda_{\,j+\half}^{\,n} \moins \gamma_{\,j+\half}^{\,n} \,\bm\, \biggr]\, u_{\,j+1}^{\,n+1} 
 \moins \biggl[\, \gamma_{\,j-\half}^{\,n} \,\bp \plus \lambda_{\,j-\half}^{\,n} \, \biggr]\, u_{\,j-1}^{\,n+1} \egal \\
 \biggl[\, 1 \moins \gamma_{\,j+\half}^{\,n} \ (\,\bp \moins \bm \,) \moins \lambda_{\,j+\half}^{\,n} \plus \lambda_{\,j-\half}^{\,n} \,\biggr]\, u_{\,j}^{\,n} 
 \plus \\ \biggl[\, \lambda_{\,j+\half}^{\,n} \moins \gamma_{\,j+\half}^{\,n} \,\bm\, \biggr]\, u_{\,j+1}^{\,n} 
 \plus \biggl[\, \gamma_{\,j-\half}^{\,n} \,\bp \plus \lambda_{\,j-\half}^{\,n} \, \biggr]\, u_{\,j-1}^{\,n} \,,
\end{multline}
The combination of IMplicit-EXplicit (\emph{IMEX)} approaches clearly appear in this formulation of Eq.~\eqref{eq:schema_CN_NL}. The major advantage over the classical \CN ~scheme is to avoid sub-iterations in the solution procedure, without loosing the accuracy and the stability.


\subsection{The \SG ~scheme}

In the nonlinear case, the \SG ~numerical schemes is written as: 
\begin{align}\label{eq:SG_NL}
c_{\,j}^{\,n} \; \frac{u_{\,j}^{\,n+1}\ -\ u_{\,j}^{\,n}}{\Delta t} 
\egal \frac{1}{\Delta x} \biggl[\, J_{\,j+\frac{1}{2}}^{\,n}\ -\ J_{\,j-\frac{1}{2}}^{\,n}   \, \biggr] \,.
\end{align}

We use the hypothesis of frozen coefficient on the interval $\big[ \, x_{j}, \, x_{j+1} \, \big]\,$. Thus, the flux at each interface $x_{\,j+\half}$ is computed with following boundary value problem:
\begin{subequations}\label{sec1_eq:SP_equation_NL}
\begin{align}
 J_{\,j+\half}^{\,n} & \egal \Pe_{\,j+\half}^{\,n} \, u \moins d_{\,j+\half}^{\,n} \, \pd{u}{x} \,, & & \forall x \ \in \ \big[ \, x_{j}, \, x_{j+1} \, \big] \,, \\
 u & \egal u_{\,j}^{\,n} \,, && x \egal x_{j} \,,\\
 u  &\egal u_{\,j+1}^{\,n} \,, && x \egal x_{j+1} \,.
\end{align}
\end{subequations}
The solution of Eq.~\eqref{sec1_eq:SP_equation_NL} is:
\begin{align*}
& J_{\,j+\half}^n  \egal \frac{1}{\Delta x} \, d_{\,j+\half}^{\,n}
\left[ \, -\ \mathcal{B}\Biggl( \, \frac{\Delta x }{d_{\,j+\half}^{\,n}} \, \Pe_{\,j+\half}^{\,n} \, \Biggr) \, u_{\,j+1}^{\,n} 
\plus \mathcal{B}\Biggl( \, - \ \frac{\Delta x }{d_{\,j+\half}^{\,n}} \, \Pe_{\,j+\half}^{\,n} \, \Biggr) u_{\,j}^{\,n} \, \right] \,.
\end{align*}

When dealing with the nonlinearities of the material properties, an interesting feature of explicit schemes is that it does not require any sub-iterations (using \textsc{Newton}--\textsc{Raphson} approach for instance). At the time iteration $n$, the material properties $c_{\,j}$, $d_{j+\half}$, $d_{j-\half}$ are \emph{explicitly} calculated at $t^{\,n}$. The CFL condition of the scheme is given by Gosse \cite{Gosse2016}: 
\begin{align}\label{eq:cfl_SG_non_linear}
\dt \cdot \max_{x \in [\, 0,\, 1\, ]} \Biggl[\, \dfrac{\Pe}{\dms} \, \tanh \left(\, \dfrac{\Pe \, \dx}{2 \, \dms} \,\right)^{-1}\, \Biggr] \ \leqslant\ \dx \cdot \frac{\cms}{\dms} \,.
\end{align}


\subsection{Numerical application}

\begin{figure}
\centering
\subfigure[a][\label{fig_AN2:cm}]{\includegraphics[scale=.6]{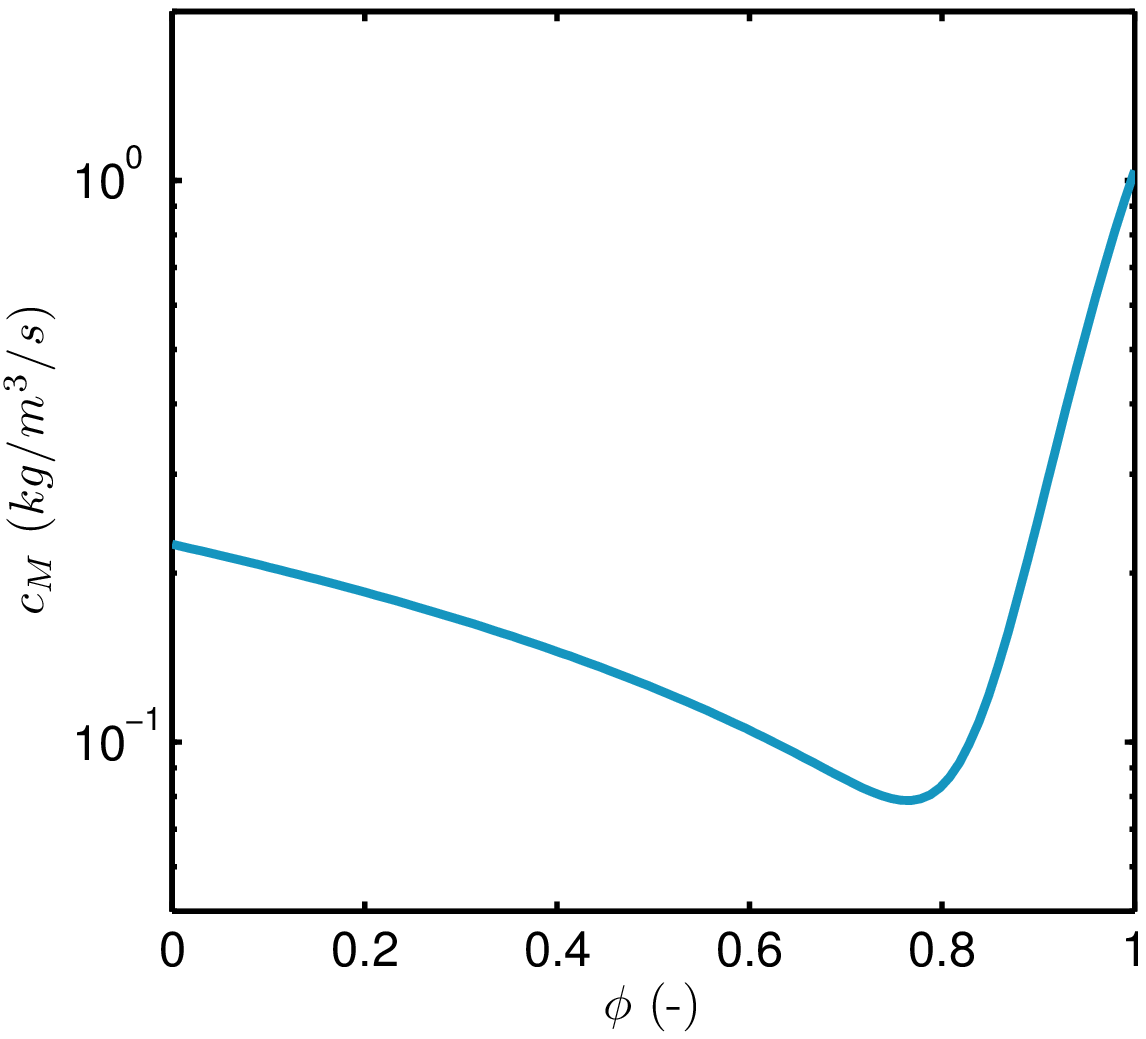}} \hspace{0.3cm}
\subfigure[b][\label{fig_AN2:dm}]{\includegraphics[scale=.6]{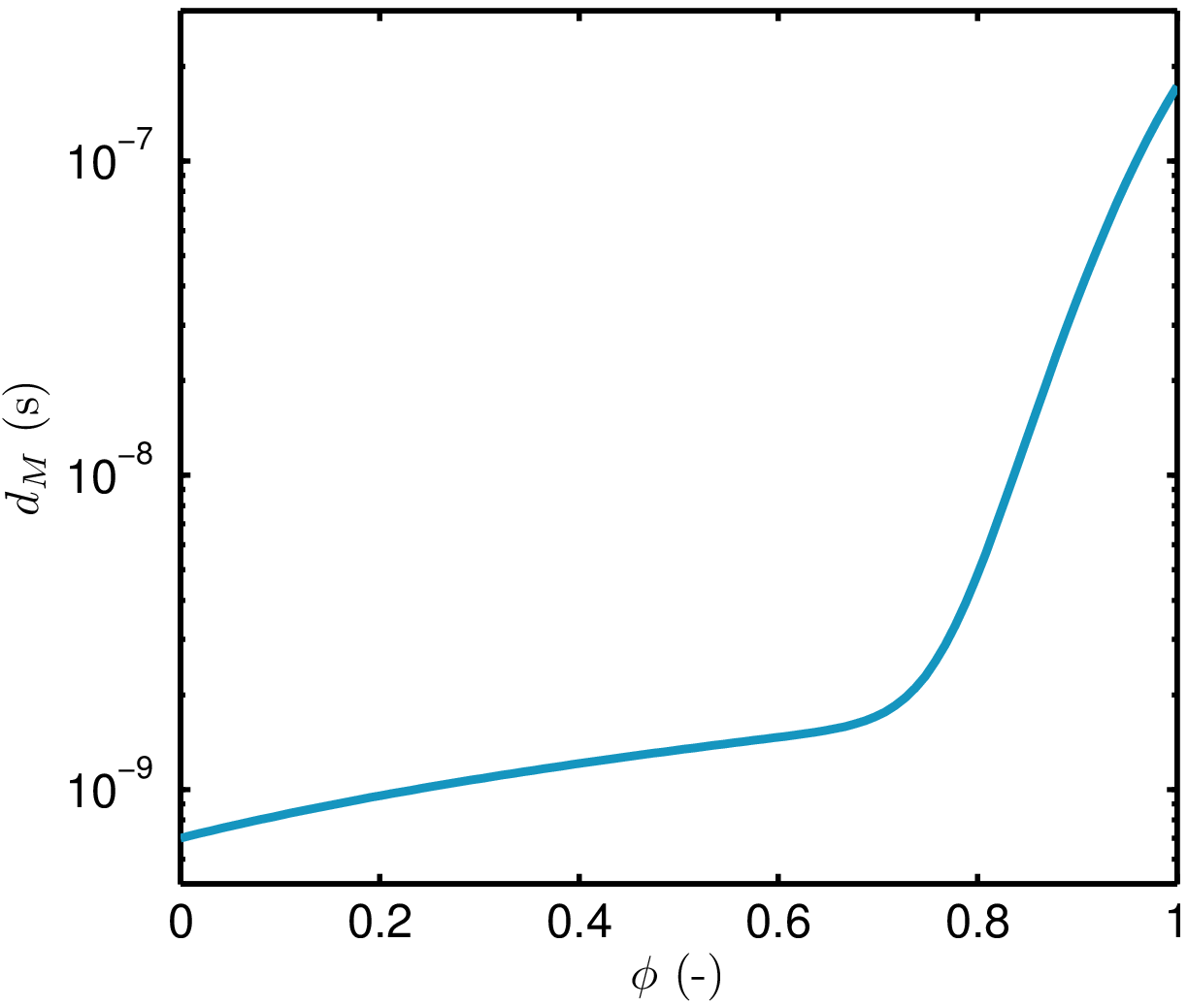}}
\caption{\small\em Variation of the moisture storage $c_M$ (a) and diffusion $d_M$ (b) as a function of the relative humidity $\phi$.} 
\end{figure}

The dimensionless properties of the materials are:
\begin{align*}
& \dms \egal 1 \plus 0.91 \, u \plus 600 \cdot \exp \biggl[ \,-10 \, \bigl(\, u \moins 2.3 \, \bigr)^2 \, \biggr] \,,\\
& \cms \egal 900 \moins 400 \, u \plus 10^4 \cdot \exp \biggl[ \,-10 \, \bigl(\, u \moins 2.3 \, \bigr)^2 \,  \biggr] \,.
\end{align*}
From a physical point of view, the storage and diffusion coefficients are given in Figures~\ref{fig_AN2:cm}~and~\ref{fig_AN2:dm}. Their variations with the relative humidity are similar to the load bearing material from \cite{Janssen2014}. The \textsc{P\'eclet} number is equal to $\Pe \egal 10$ corresponding to a mass average velocity of $0.01$ $\mathsf{m/s}\,$. A first order approximation of the air transfer in porous material is given by \cite{Belleudy2016}: 
\begin{align}\label{eq:approx_vitesse}
  \mathsf{v} \egal - \, \frac{\kappa_{\,m}}{\mu_{\,a}} \, \pd{P}{x} \,,
\end{align}
where $\kappa_{\,m}$ is the air permeability of the material, $\mu_{\,a} \egal 1.8 \cdot 10^{\,-5}$ \unite{Pa.s} is the dynamic viscosity of air and $P$ is the  air pressure. Therefore, such condition can be obtained for a material with an air permeability of the order of $\kappa_{\,m} \egal 10^{\,-9} \ \mathsf{m^{\,2}}$, with an air pressure difference, between both sides of the material, of $10 \ \mathsf{Pa}\,$. These conditions easily occur in buildings. Moreover, a number of hygroscopic building materials as cellulose, wood fibre or hemp concrete, has an air permeability of such order \cite{KumarKumaran1996, ASHRAE2013}.

\begin{figure}
\centering
\includegraphics[width=0.65\textwidth]{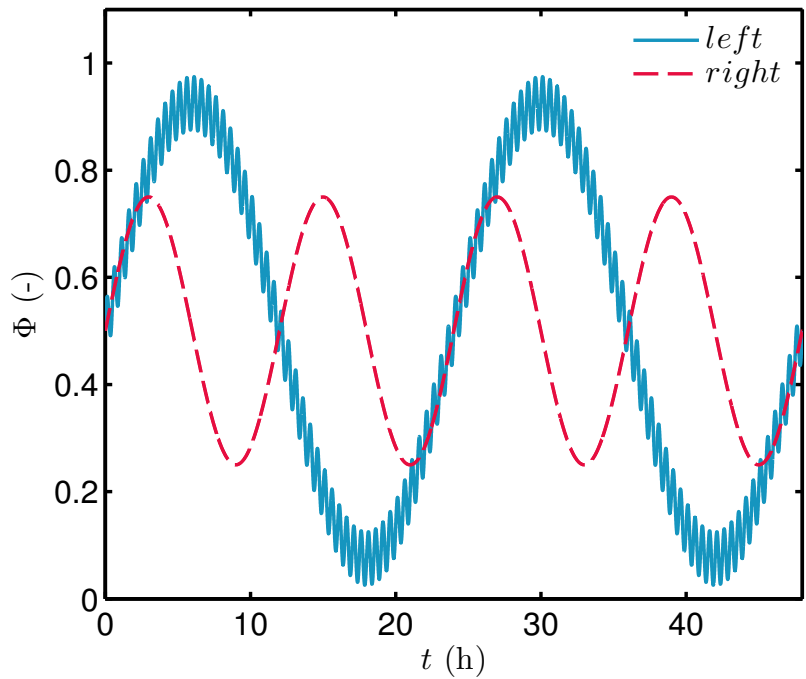}
\caption{\small\em Boundary conditions.}
\label{fig_AN2:BC}
\end{figure}

The initial vapour pressure is uniform $\Pvi \egal 1.16 \dix{3}$  $\mathsf{Pa}$, corresponding to a relative humidity $\phi \egal 50$ $\mathsf{\%}$. No liquid flow is taken into account at the boundaries. The \textsc{Biot} numbers are fixed to $\BivL \egal 28.75$ and $\BivR \egal 4.28 \,$. The ambient vapour pressure at the boundaries are different from the previous case study, $\uR$ and $\uL$ follow sinusoidal variations from the dry to saturate state, forcing periodical conditions: 
\begin{align*}
& \uR \egal 1 \plus 0.85 \; \sin \left(\, \frac{2\pi}{24} \, \ts \,\right) \plus 0.1 \; \sin \left(\, 4\pi \, \ts \,\right) \,, \\
& \uL \egal 1 \plus 0.5 \; \sin \left(\, \frac{2\pi}{12} \, \ts \,\right) \,.
\end{align*}
The physical boundary conditions are illustrated in Figure~\ref{fig_AN2:BC}. The material is thus excited until the capillary state. The final simulation time is fixed to $\tau^{\,\star} \egal 48 \,$.

\begin{figure}
\centering
\subfigure[a][\label{fig_AN2:time_evolution}]{\includegraphics[scale=.6]{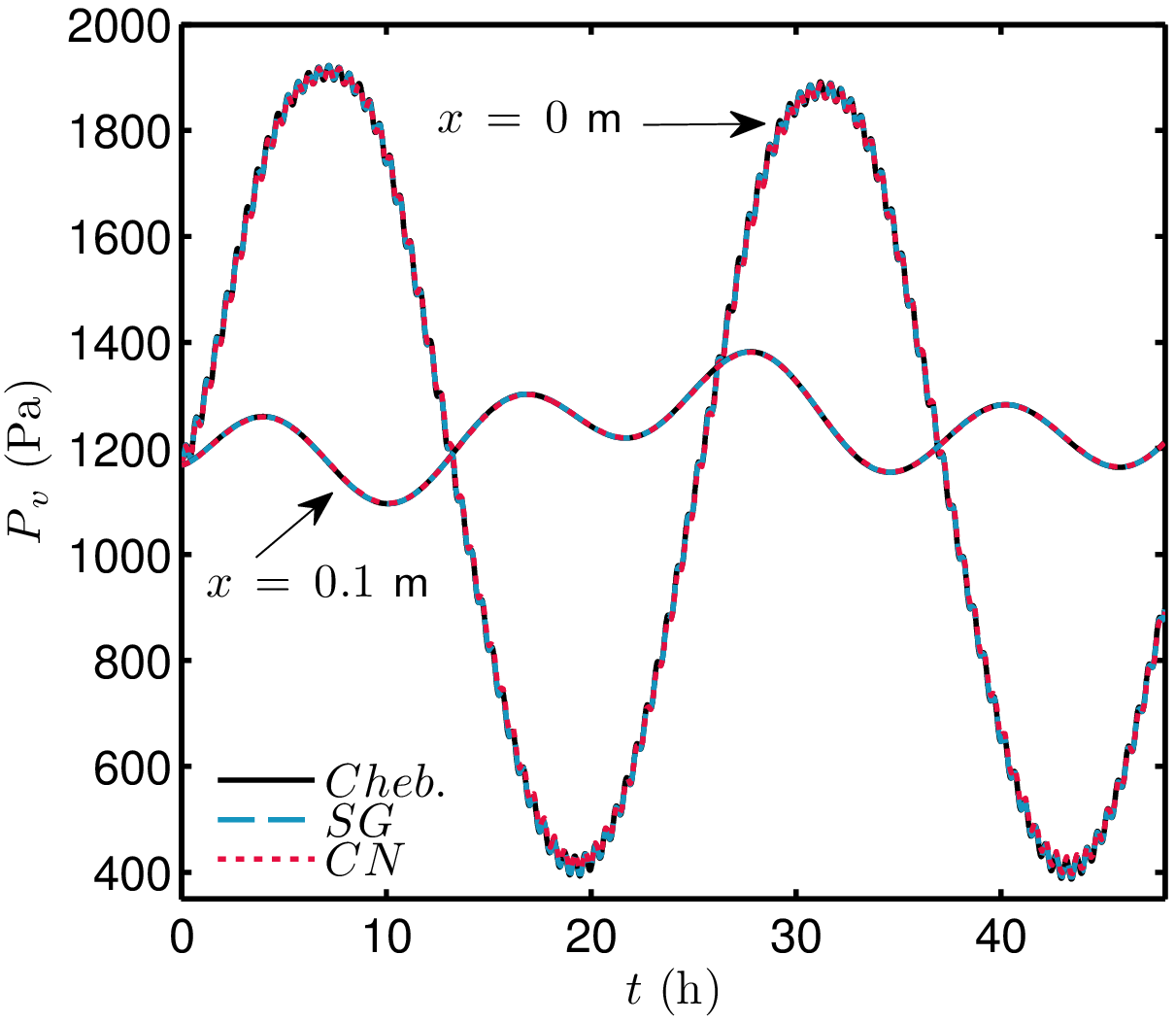}} \hspace{0.3cm}
\subfigure[b][\label{fig_AN2:profil}]{\includegraphics[scale=.6]{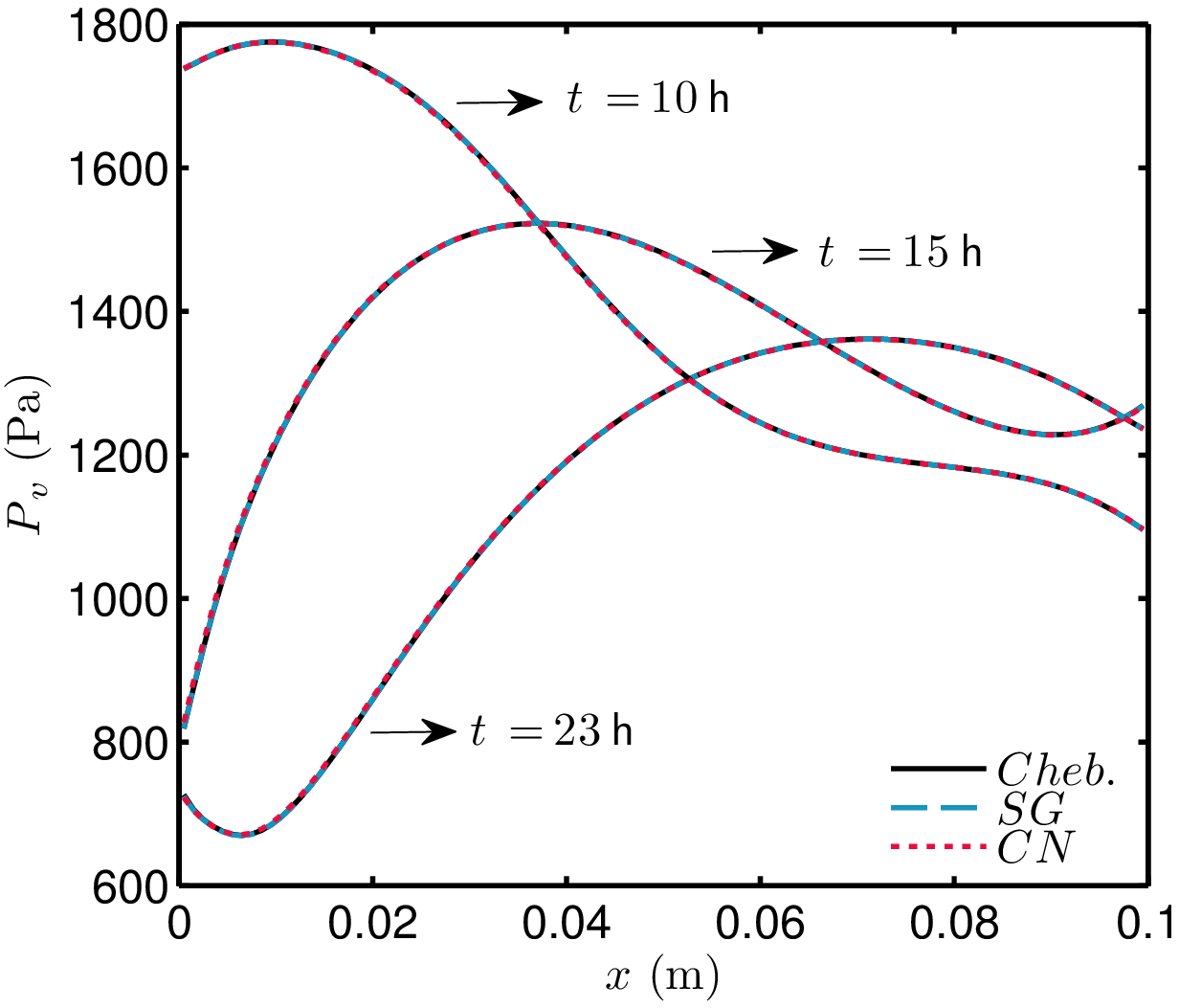}}
\caption{\small\em Vapour pressure time evolution at $x \in \{0,\,0.1 \}$ $\mathsf{m}$ (a) and profiles for $t \in \{10,\,15,\, 23\}$ $\mathsf{h}$ (b).} 
\end{figure}

The solution of the problem has been computed with following discretisation parameters: $\dt^{\,\star} \egal 1 \cdot 10^{-3}$ and $\dx^{\,\star} \egal 1 \cdot 10^{-2}\,$. The time variation of the vapour pressure on the two extremities of the building component is given in Figure~\ref{fig_AN2:time_evolution}. The vapour pressure is increasing according to the variation at the left boundary condition, which has a higher \textsc{Biot} number and a higher amplitude. The diffusion process can be observed going from the left to the right boundary. The diffusion process can also be seen on the three profiles of the vapour pressure illustrated in Figure~\ref{fig_AN2:profil}.

\begin{figure}
\centering
\subfigure[a][\label{fig_AN2:err_L2_fdt_case1}]{\includegraphics[scale=.6]{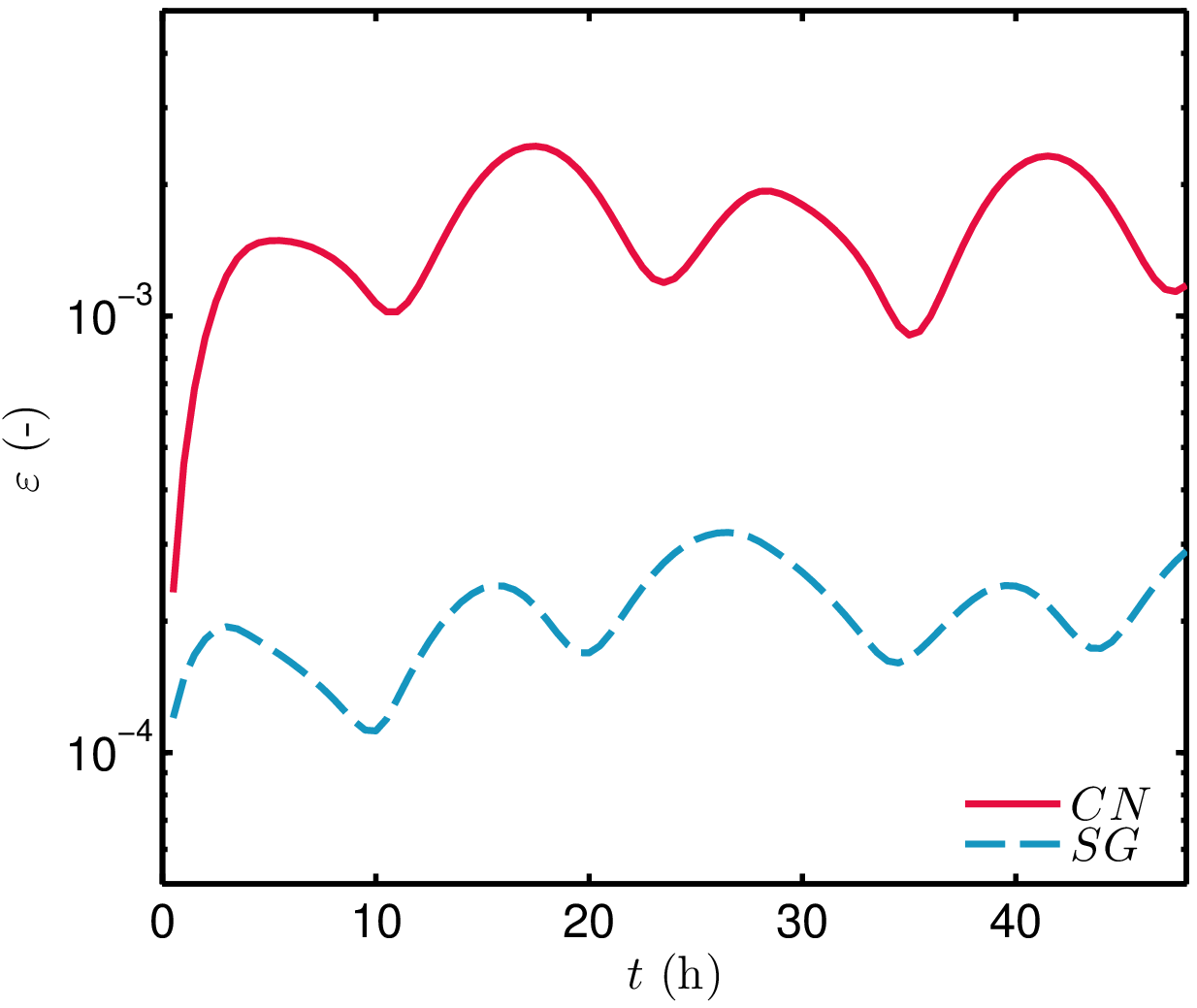}} \hspace{0.3cm}
\subfigure[b][\label{fig_AN2:err_L2_fdx_case1}]{\includegraphics[scale=.6]{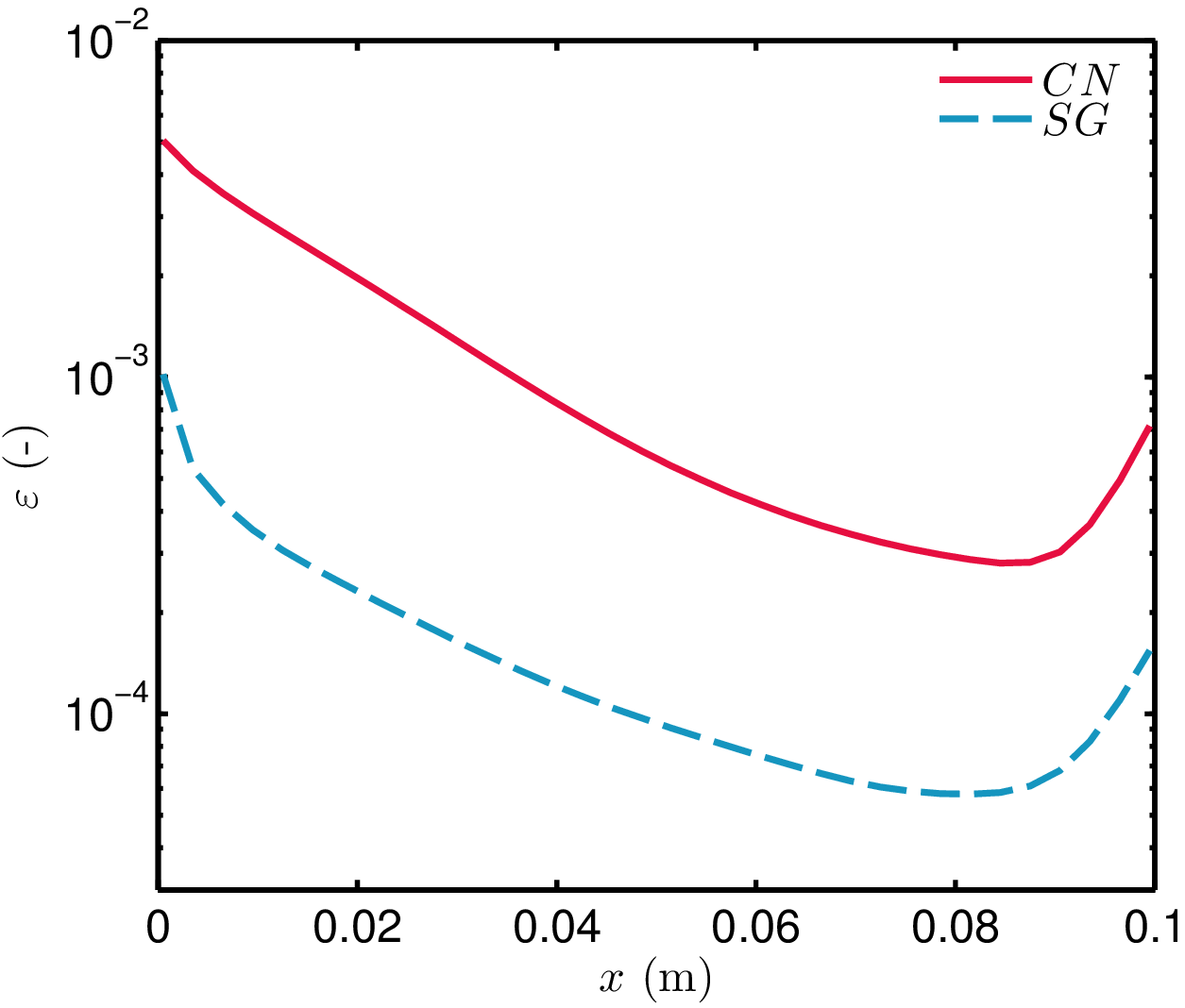}}
\caption{\small\em $\mathcal{L}_{\,2}$ error as a function of time $t$ (a) and as a function of space $x$ (b).} 
\end{figure}

Both schemes succeed in representing the advection-diffusion phenomena in the building component. The $\mathcal{L}_{\,2}$ error of the \CN ~and of the \SG ~schemes are presented in Figure~\ref{fig_AN2:err_L2_fdt_case1} and in Figure~\ref{fig_AN2:err_L2_fdx_case1}, one in function of time and the other in function of the space. The \SG ~scheme has shown to be more accurate than the \CN ~one. For a better understanding of the accuracy of the schemes a convergence study is presented.

\begin{figure}
\centering
\subfigure[a][\label{fig_AN2:err_L2_fdt}]{\includegraphics[scale=.6]{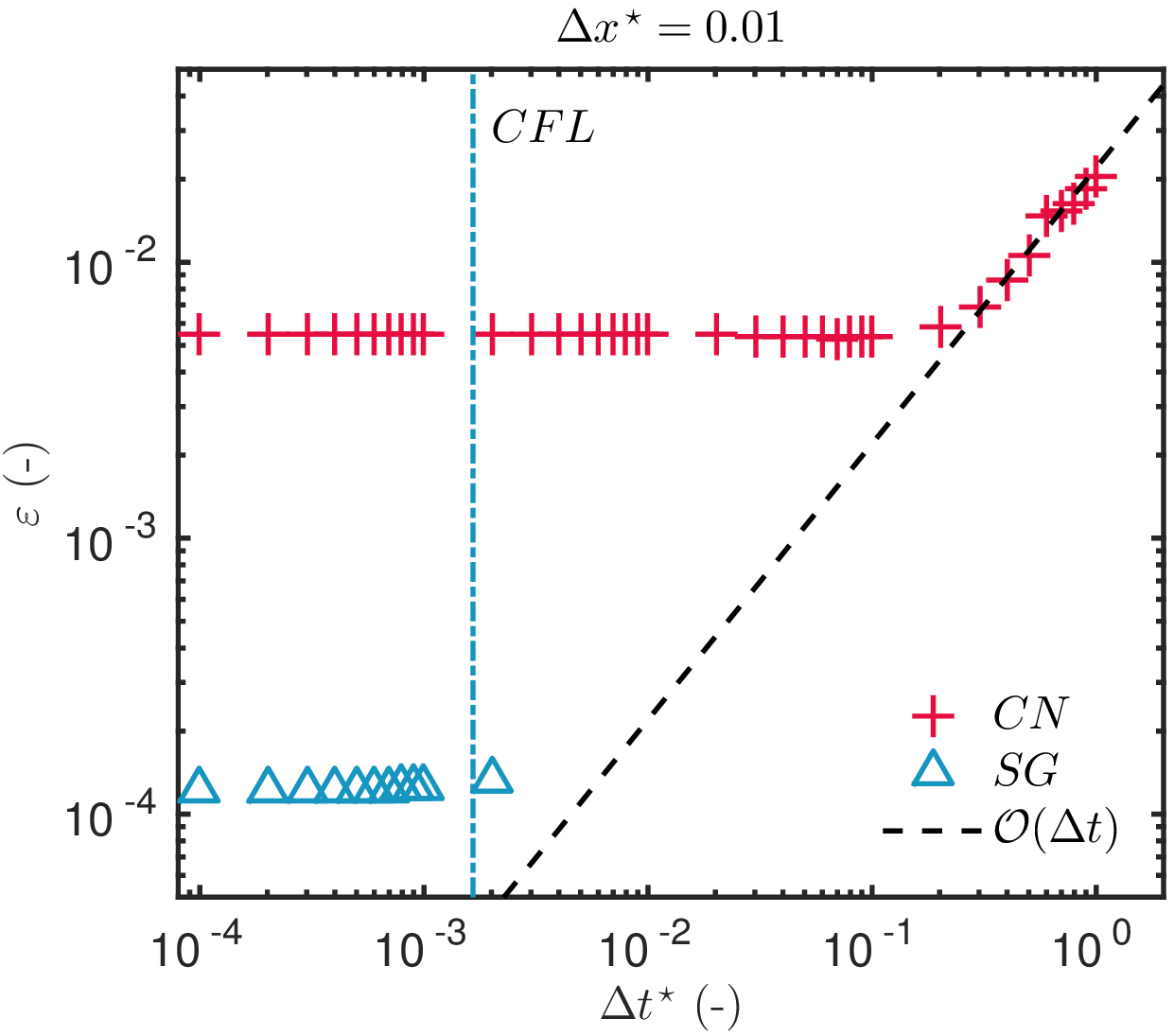}} \hspace{0.3cm}
\subfigure[b][\label{fig_AN2:err_L2_fdx}]{\includegraphics[scale=.6]{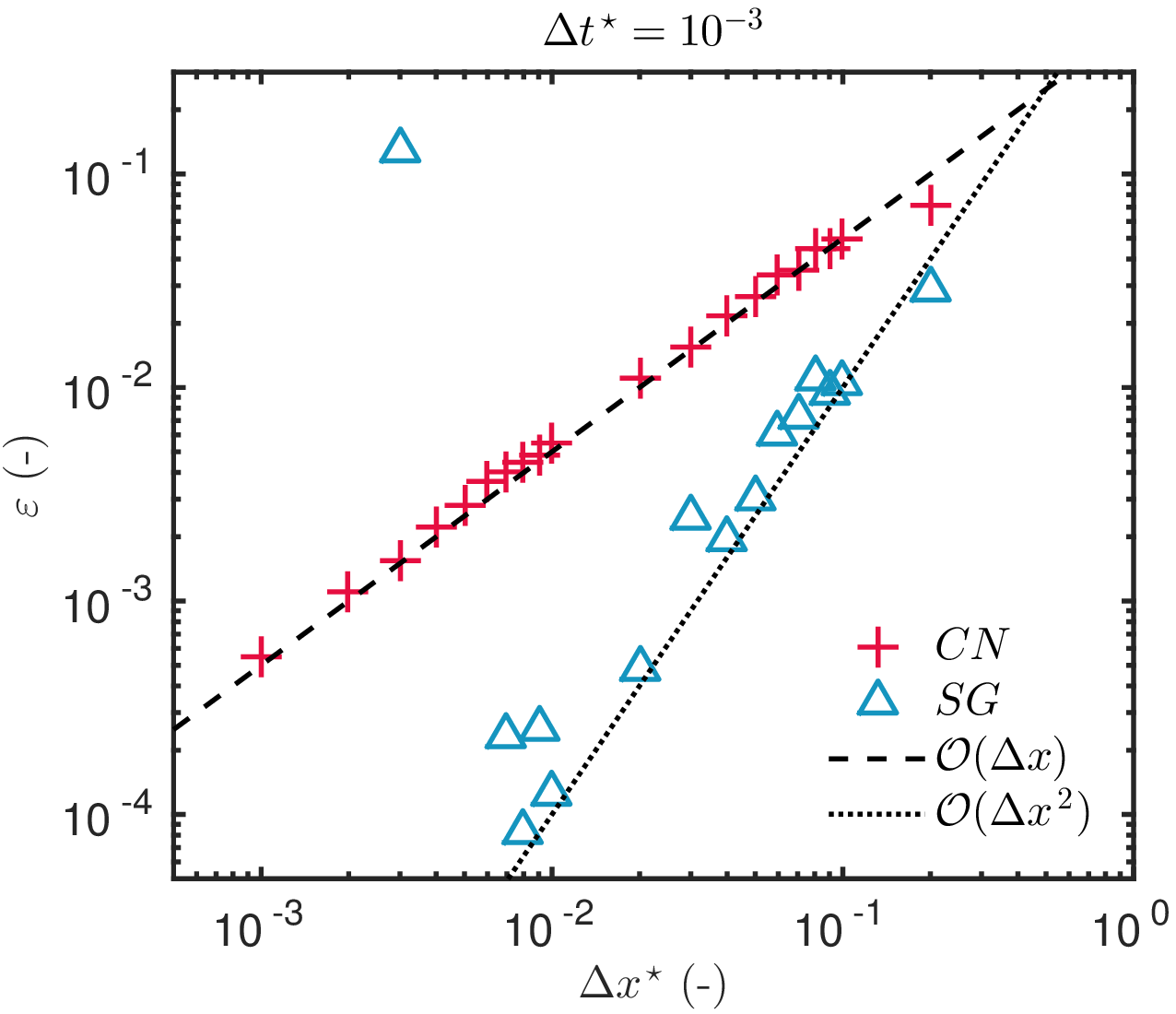}}
\caption{\small\em $\mathcal{L}_{\,2}$ error $\varepsilon$ for a fixed $\Delta x^{\,\star} \egal 10^{\,-2}$ as a function of $\Delta t^{\,\star}$ (a), and $\mathcal{L}_{\,2}$ error $\varepsilon$ for a fixed $\Delta t^{\,\star} \egal 10^{\,-3}$ as a function of $\Delta x^{\,\star}\,$.}
\end{figure}

The problem is then computed with $\dx^{\,\star} \egal 10^{\,-2}$ and for different values of $\dt^{\,\star}\,$. For each value of $\dt^{\,\star}\,$, the $L_{\,2}$ error was computed between the solutions of the schemes and a \emph{Chebfun} reference solution. Figure~\ref{fig_AN2:err_L2_fdt} shows the results of this convergence study. Meanwhile in Figure~\ref{fig_AN2:err_L2_fdx} the $\dt^{\,\star}$ is fixed to $10^{\,-3}$ and the $\mathcal{L}_{\,2}$ error is computed for different values of $\dx^{\,\star}\,$. The CFL condition of the \SG ~scheme is given by Eq.~\eqref{eq:cfl_SG_non_linear}, with $\dt^{\,\star} \ \leqslant \ 1 \dix{-3}$. The \SG ~scheme can only be computed when the CFL condition is respected, while the modified \CN ~scheme is unconditionally stable (at least for linear problems). Despite of the stability of the \mCN ~scheme, the \SG ~approach gains in terms of accuracy. Moreover, the stability of the \mCN ~scheme does not neccessary imply an accurate solution \cite{Patankar1980}. The choice of the discretisation depends on the boundary condition as well as the characteristic diffusion time in the material, to accurately represent the physical phenomenon. Figure~\ref{fig_AN2:err_L2_fdt} reveals that the \mCN ~is first order of accuracy in time $\O(\dt)\,$. The modification of the classical \CN ~scheme implies loosing the $\O(\, \dt ^{\,2})$ accuracy in an effort to avoid the sub-iterations due to nonlinearities. For the \SG ~scheme, it is not possible to verify the order of accuracy in time. Indeed, before the CFL condition, the error of the scheme is influenced by the order of $\dx^{\,\star}\,$. In Figure~\ref{fig_AN2:err_L2_fdx} the order of accuracy regarding the space discretisation, corresponds to $\O(\dx^{\,2})$ for \SG ~and $\O(\, \dx)$ for the \mCN. Besides the CFL condition, the \SG ~scheme has a minimum restriction regarding to $\dx^{\,\star}\,$. If the value of $\dx^{\,\star}$ is below the limit, the solution diverges.

An interesting advantage of the \SG ~scheme compared to the \CN ~approach, is the ease of implementation. The algorithm is written using an explicit formulation and the fluxes are calculated analytically for each mesh element. It may allow to achieve almost perfect scaling on high-performance computer systems \cite{Chetverushkin2012}.

\begin{figure}
\centering
\subfigure[a][\label{fig_AN2:cpu_time_fdt}]{\includegraphics[width=0.485\textwidth]{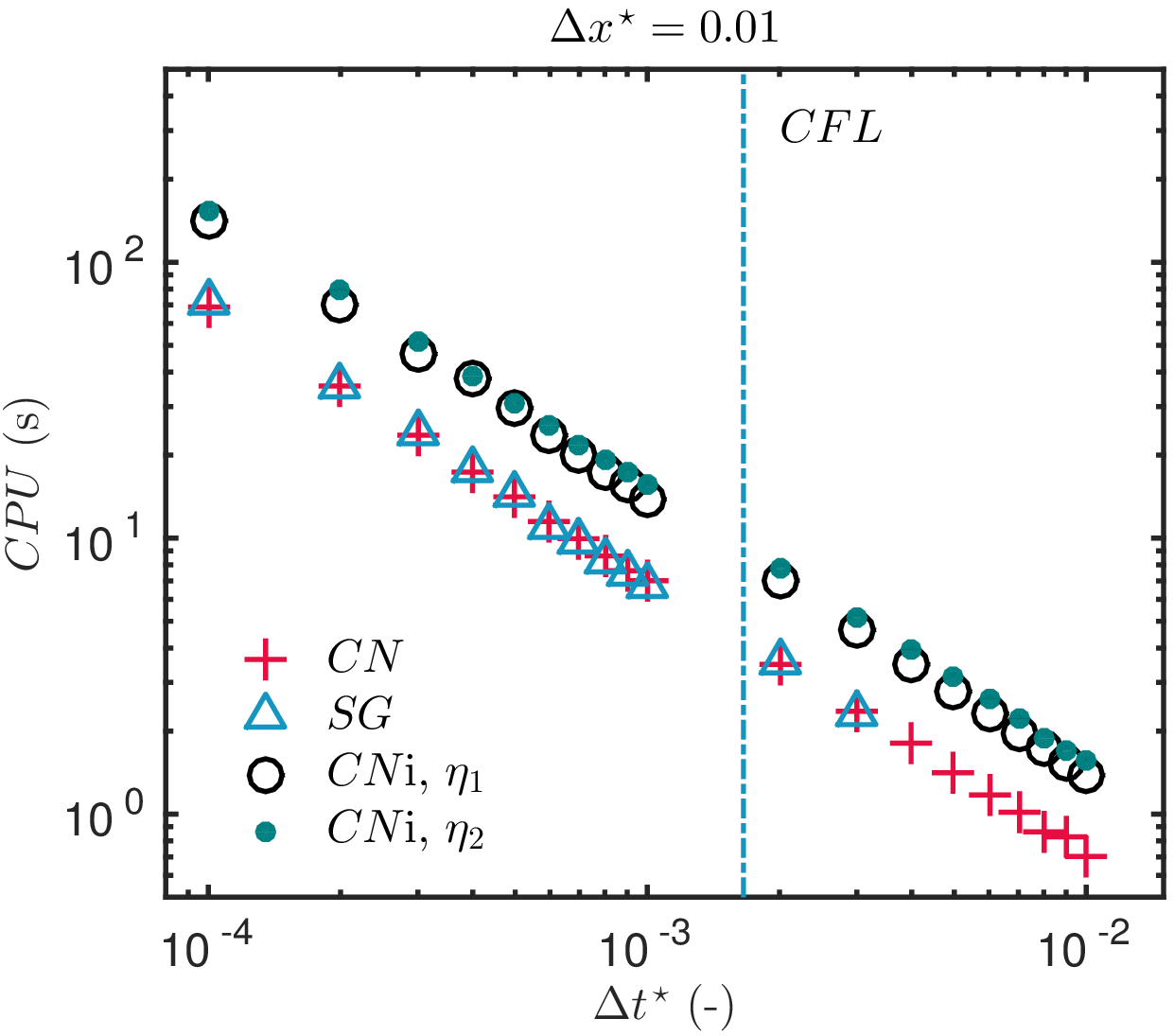}}
\subfigure[b][\label{fig_AN2:cpu_time_fdx}]{\includegraphics[width=0.485\textwidth]{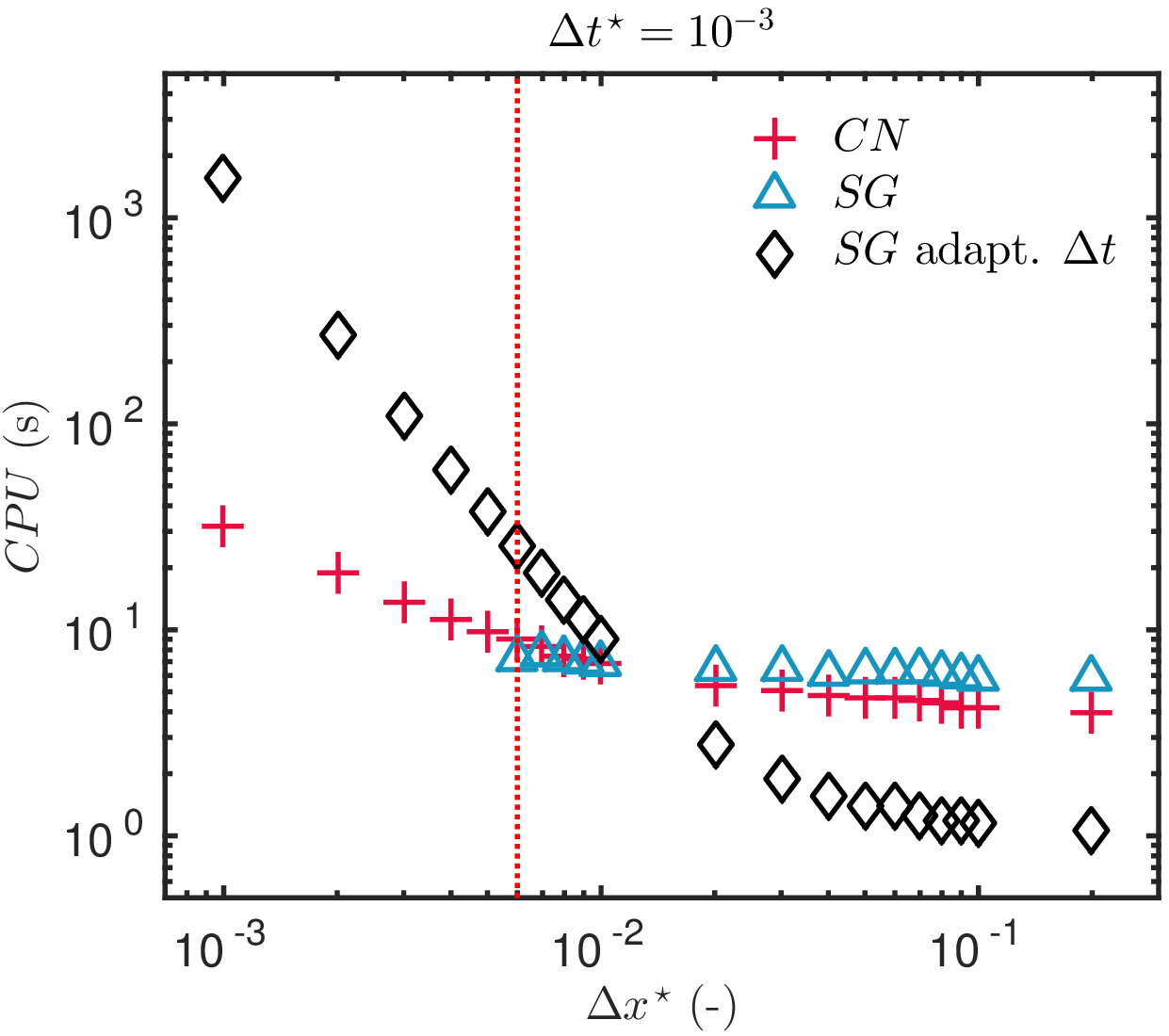}}
\caption{\small\em Computational time for $\Delta x^{\,\star} \egal 10^{\,-2}$ as a function of $\Delta t^{\,\star}$ (a), and for $\Delta t^{\,\star} \egal 10^{\,-3}$ as a function of $\Delta x^{\,\star}\,$.}
\end{figure}

For this numerical application, the CPU time of each numerical scheme has been evaluated using \texttt{Matlab} platform on a computer using Intel i7 CPU and 32GB of RAM. Results are presented in Figure~\ref{fig_AN2:cpu_time_fdt} as a function of $\dx^{\,\star}$ and in Figure~\ref{fig_AN2:cpu_time_fdx} as function of $\dt^{\,\star}$. The \CN ~and \SG ~schemes have the same order of magnitude of CPU time. On the other hand, the \SG ~scheme has been implemented with an adaptive $\dt^{\,\star}$ using \texttt{Matlab} function \texttt{ode113}. As illustrated in Figure~\ref{fig_AN2:cpu_time_fdx}, it enables to gain a significant computational time when $\dx$ is relatively large, whithout loosing any accuracy. The classical \CN ~scheme has also been used to compute a problem solution, using a tolerance $\eta_{\,1}  \egal \dt^{\,\star} $ and $\eta_{\,2} \egal 0.5 \, \dt^{\,\star}$ for the convergence of the sub-iterations required to treat the nonlinearities. Figure~\ref{fig_AN2:cpu_time_fdt} shows that the CPU time is almost multiplied by two using this approach. Thus, the use of the classical \CN ~scheme has strong disadvantages compared to the two other proposed schemes. The \mCN ~and \SG ~approaches enable to compute an accurate solution two times faster.

Table~\ref{tab:synthesis_num_scheme} summarises the features of both numerical schemes investigated, highlighting the efficiency of the \SG ~approach to solve the advective-diffusive moisture equation.

\begin{table}
\centering
\begin{tabular}{@{}m{.75\textwidth}}
\hline
\hline
\textbf{\SG} \\
\hline
$\bullet$ Well balanced \\
$\bullet$ Asymptotic preserving  \\
$\bullet$ Explicit form of the solution, no sub-iterations required to treat the non-linearities  \\
$\bullet$ CFL stability condition scaling with $\dx$ for large spatial grid  \\
$\bullet$ Reduced CPU with an adaptative time step algorithm  \\
$\bullet$ Lower absolute accuracy  \\
\hline
\textbf{\mCN} \\
\hline
$\bullet$ Unconditionally stable (at least for linear problems) \\
$\bullet$ Explicit expression of the material properties, no sub-iterations required to treat the non-linearities  \\
\hline
\hline
\end{tabular}
\bigskip
\caption{\small\em Synthesis of the numerical schemes features.}
\label{tab:synthesis_num_scheme}
\end{table}


\section{Comparing numerical results with experimental data}

Previous sections aimed at illustrating the advantages of the \SG ~scheme to compute the problem of moisture transfer by diffusion and advection mechanisms through a porous material, which can be used to reduce the discrepancies between experimental and numerical results. For this, experimental data from \cite{James2010} are used for the comparison, which considered a gypsum board initially conditioned at $\phi \egal 0.3$, submitted to a $48$ $\mathsf{h}$ adsorption-desorption cycle (30--72--30). A constant surface transfer coefficient is assumed equal to $2.41 \dix{-8}$~$\mathsf{s/m}$ and the material properties are given in Table~2 from \cite{James2010}. The boundary conditions used in the computation are:
\begin{align*}
\dm \, \pd{\Pv}{x} \moins \Pe \, \Pv &\egal 
\hvL \cdot \left( \, \Pv \moins \PvL \, \right) \, , && x \egal 0 \,, \\[3pt]
 - \, \dm \, \pd{\Pv}{x} \plus \Pe \, \Pv &\egal 
\hvR \cdot \left( \, \Pv \moins \PvR \, \right)\, ,&& x \egal L \,.
\end{align*}

\begin{figure}
\centering
\subfigure[a][\label{fig_exp:Rh_ft_x1}]{\includegraphics[width=0.48\textwidth]{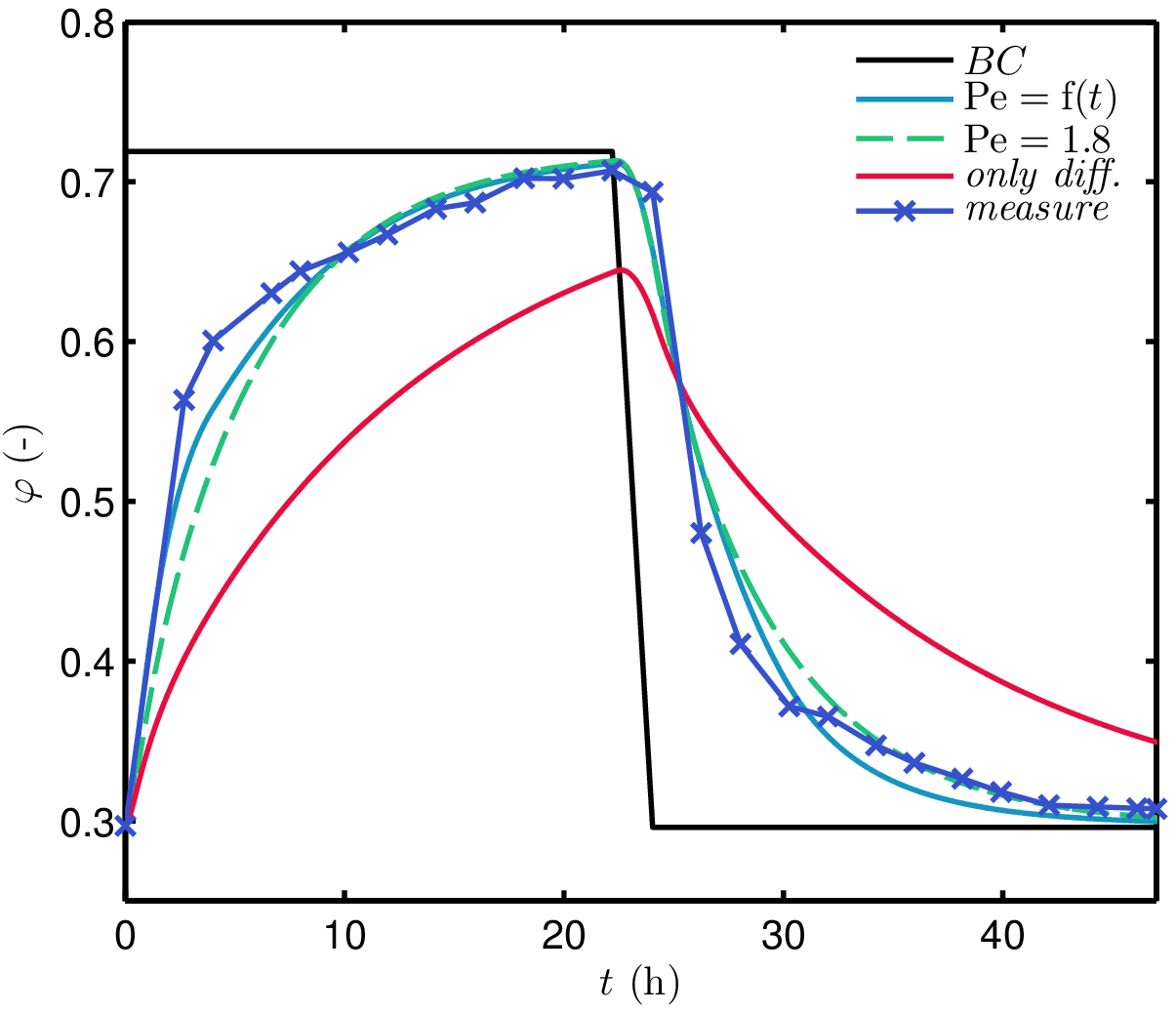}} \hspace{0.3cm}
\subfigure[b][\label{fig_exp:Rh_ft_x2}]{\includegraphics[width=0.48\textwidth]{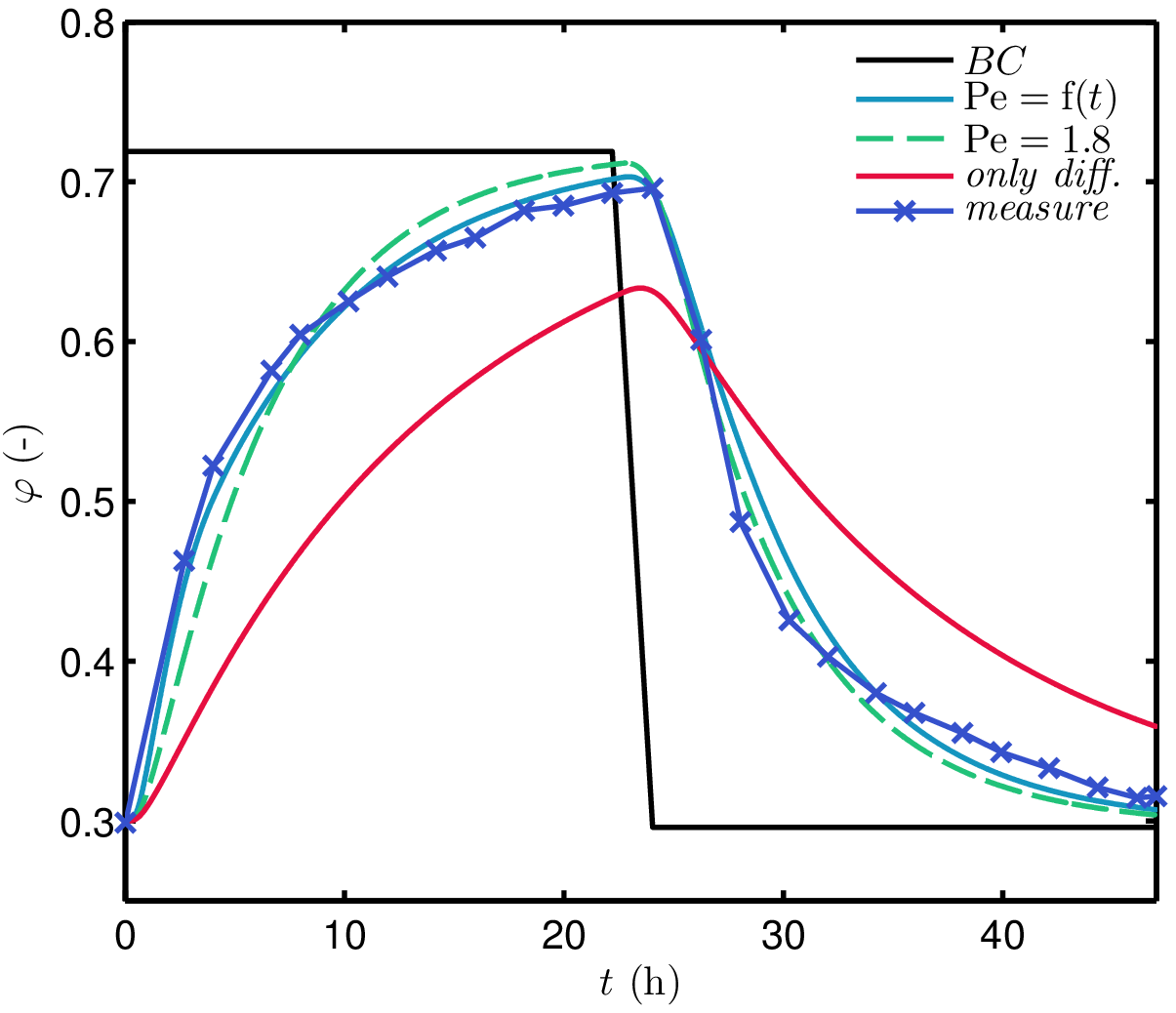}}
\caption{\small\em Measured and simulated relative humidity at $x \egal 12.5$ $\mathsf{mm}$ (a) and $x \egal 25$ $\mathsf{mm}$ (b).}
\end{figure}

The \SG ~numerical scheme has been used to compute the solution of the moisture transfer in the material, considering $\dx^{\,\star} \egal 10^{\,-2}$ and an adaptative time step. It should be noted that the \CN ~approach could have been used to compute the solution. It would also provide an accurate solution to analyse the physical phenomena. Nevertheless, given all the advantages reported in Table~\ref{tab:synthesis_num_scheme}, the \SG ~scheme was choosen for this comparison. Figures~\ref{fig_exp:Rh_ft_x1}~and~\ref{fig_exp:Rh_ft_x2} show the evolution of the relative humidity computed with a constant \textsc{P\'eclet} number and with a model considering only moisture diffusion. The \textsc{P\'eclet} number has been estimated as $\Pe \egal 1.8\,$, ensuring a $\mathcal{L}_2$ error with measurement $\varepsilon \egal 5 \cdot 10^{\,-3}\,$. The moisture diffusion model provides the same trends as the results from \cite{James2010} but does not represent well the physical phenomenon since it ignores the important advective contribution to the moisture transfer. The numerical results from the convection model do not underestimate the adsorption process or overestimate the desorption process, contrarily to the model that only considers the diffusion as a transport mechanism.

Remembering that $\Pe \egal \dfrac{\mathsf{v} \cdot L}{\Rv \cdot T \cdot \dmref}\,$, the \textsc{P\'eclet} number has been estimated as $\Pe \egal 1.8$. We assume the temperature in the material is $23$ \unite{^\circ C}, the reference moisture transport coefficient is $\dms \egal 5.6 \cdot 10^{\,-11}$ \unite{s} and the length of the material is $L \egal 37.5$ \unite{mm}. The mass average velocity in the material is then $\mathsf{v} \egal 0.4 $ \unite{mm/s}. On the other hand, the mass average velocity in the material is given by Eq.~\eqref{eq:approx_vitesse} where $\kappa_{\,m} \egal 4 \cdot 10^{\,-9}$ \unite{m^2} is the air permeability of the material \cite{KumarKumaran1996}. Thus, at the point of observation $x \egal 12.5$ \unite{mm}, the estimated velocity is due to a difference of air pressure of $\Delta \Pv \egal 0.02$ \unite{Pa}, which is reasonable considering the experimental set-up.

\begin{figure}
\centering
\subfigure[a][\label{fig_exp:XPv_ft_x1}]{\includegraphics[width=0.48\textwidth]{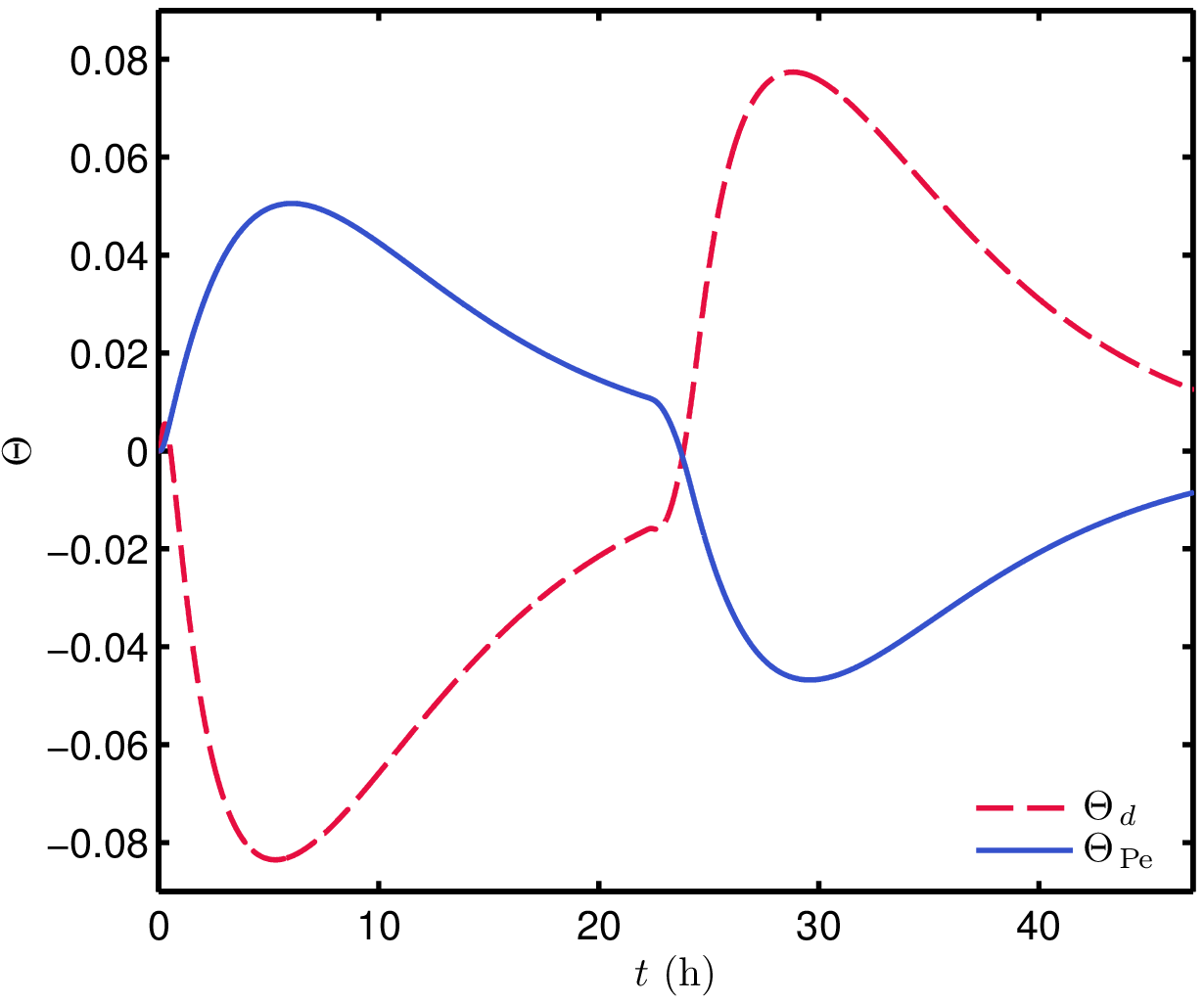}} \hspace{0.3cm}
\subfigure[b][\label{fig_exp:XPv_ft_x2}]{\includegraphics[width=0.48\textwidth]{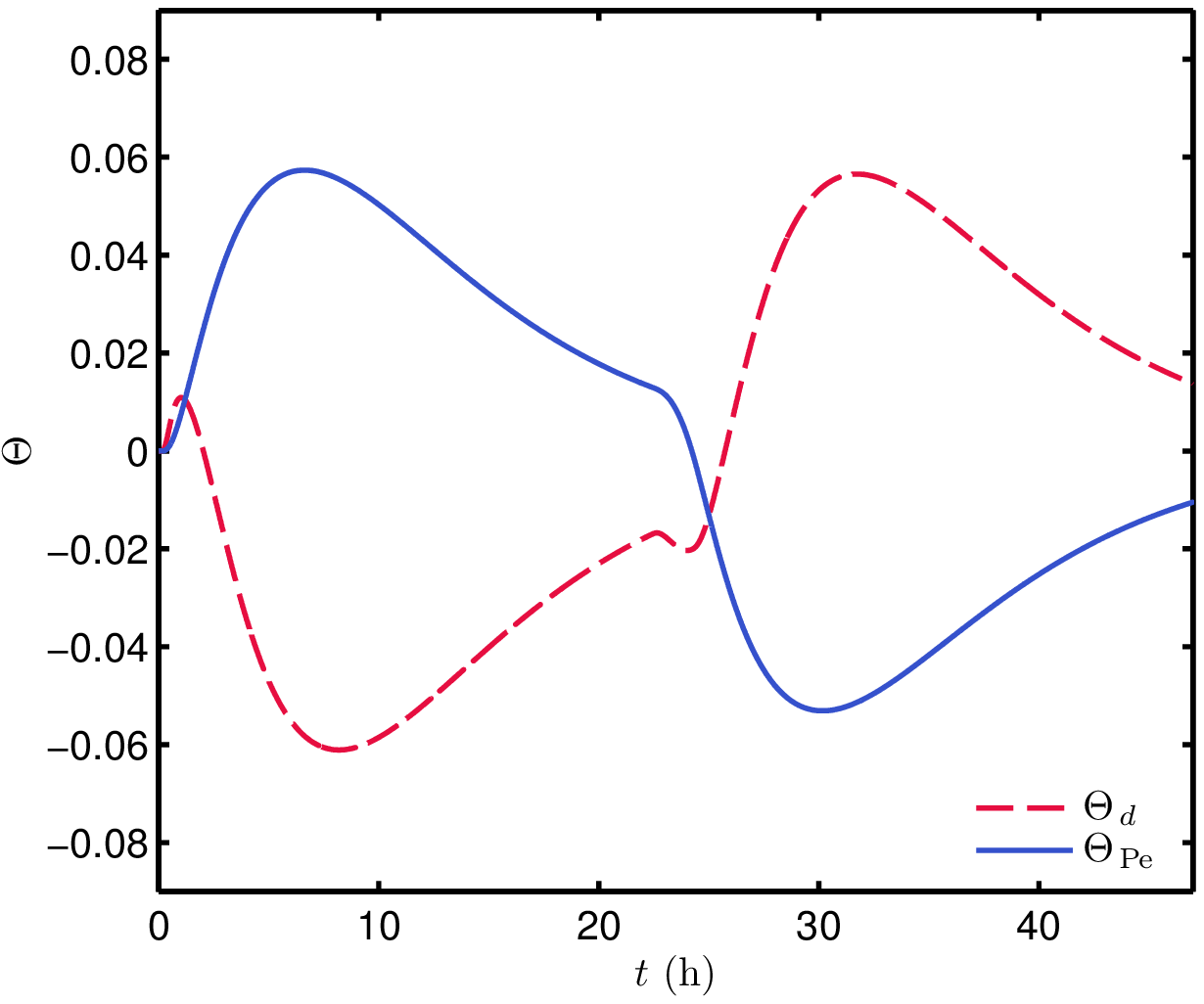}}
\caption{\small\em Sensitivity coefficients of parameters $\dms$ and $\Pe$ at $x \egal 12.5$ $\mathsf{mm}$ (a) and $x \egal 25$ $\mathsf{mm}$ (b).}
\end{figure}

In order to compare the relative importance of the terms of advective-diffusive moisture governing equation (Eq.~\eqref{eq:moisture_equation_1D}), a brief and local sensitivity analysis is carried out by defining the following sensitivity functions $\Theta$:
\begin{align*}
& \Theta_{\,d} \egal \dms \, \pd{\varphi}{\dms} \,, && \Theta_{\,\Pe} \egal \Pe \, \pd{\varphi}{\Pe} \,.
\end{align*}
Each sensitivity function $\Theta$ evaluates the sensitivity of the numerically computed field $\varphi$ with respect to parameter $\dms$ and $\Pe$. A small magnitude value of $\Theta$ indicates large changes in the parameter yield small changes in $\varphi$. Figures~\ref{fig_exp:XPv_ft_x1} and \ref{fig_exp:XPv_ft_x2} provides the time evolution of the sensitivity $\Theta$ for each parameter, at both measurement points. The sensitivity increases at the moment corresponding to the transient regimes of the simulation ($t \in \bigl[\,0, \, \,10 \bigr] \cup \bigl[\,24, \, \,34 \bigr] $ \unite{h}). Then, it decreases as the simulation reach the steady state regime. At $x \egal 12.5$ $\mathsf{mm}$, the relative humidity is more sensible to $\dms$ than $\Pe$. At $x \egal 25$ $\mathsf{mm}$, both parameters have the same order of magnitude of sensitivity. This local sensitivity analysis illustrates the importance of considering the moisture advection transfer for this material and for the relative humidity range used in the experiments from \cite{James2010}.

\begin{figure}
\centering
\subfigure[a][\label{fig_exp:1_T_ft}]{\includegraphics[width=0.48\textwidth]{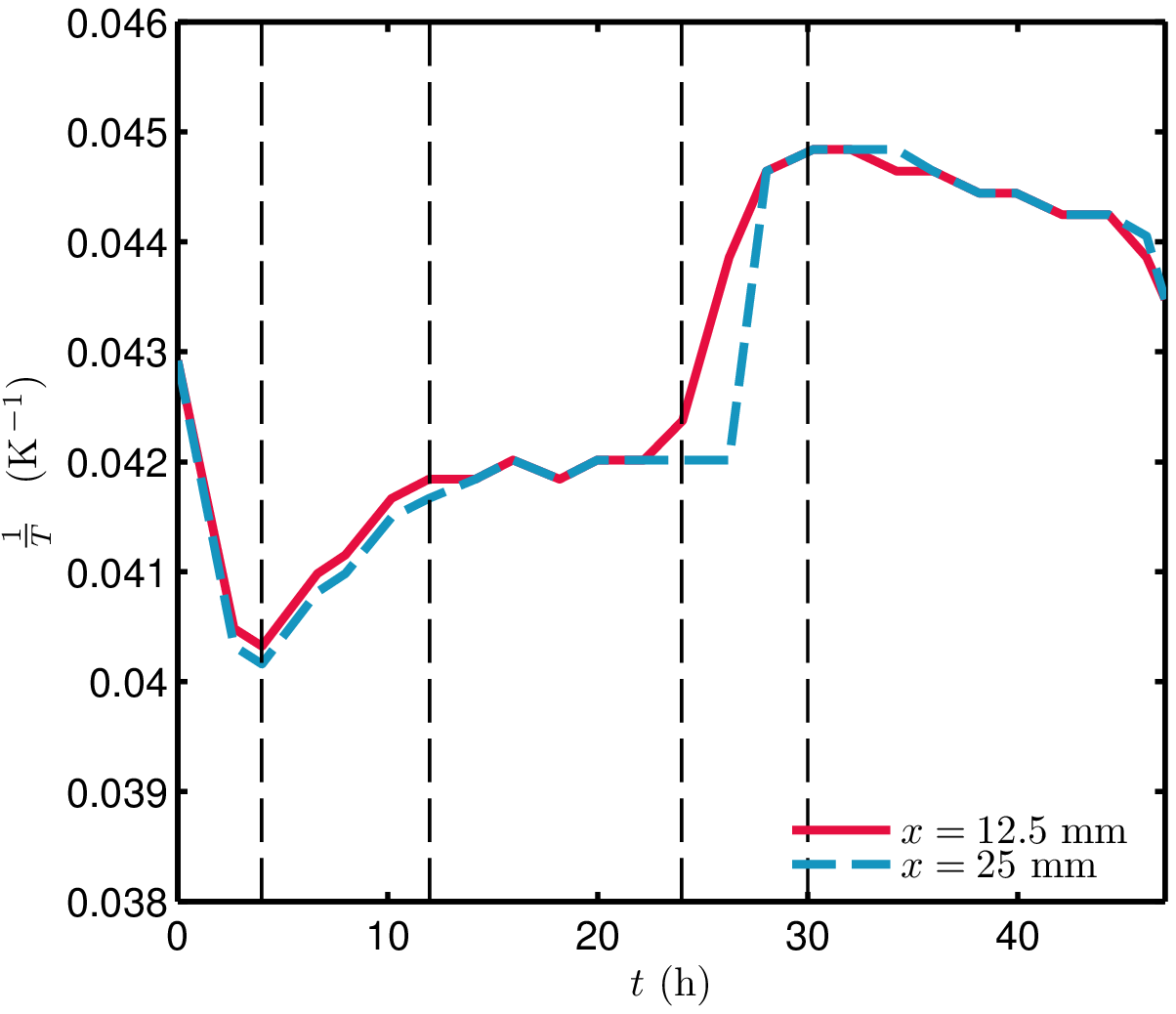}}
\subfigure[b][\label{fig_exp:Pe_ft}]{\includegraphics[width=0.48\textwidth]{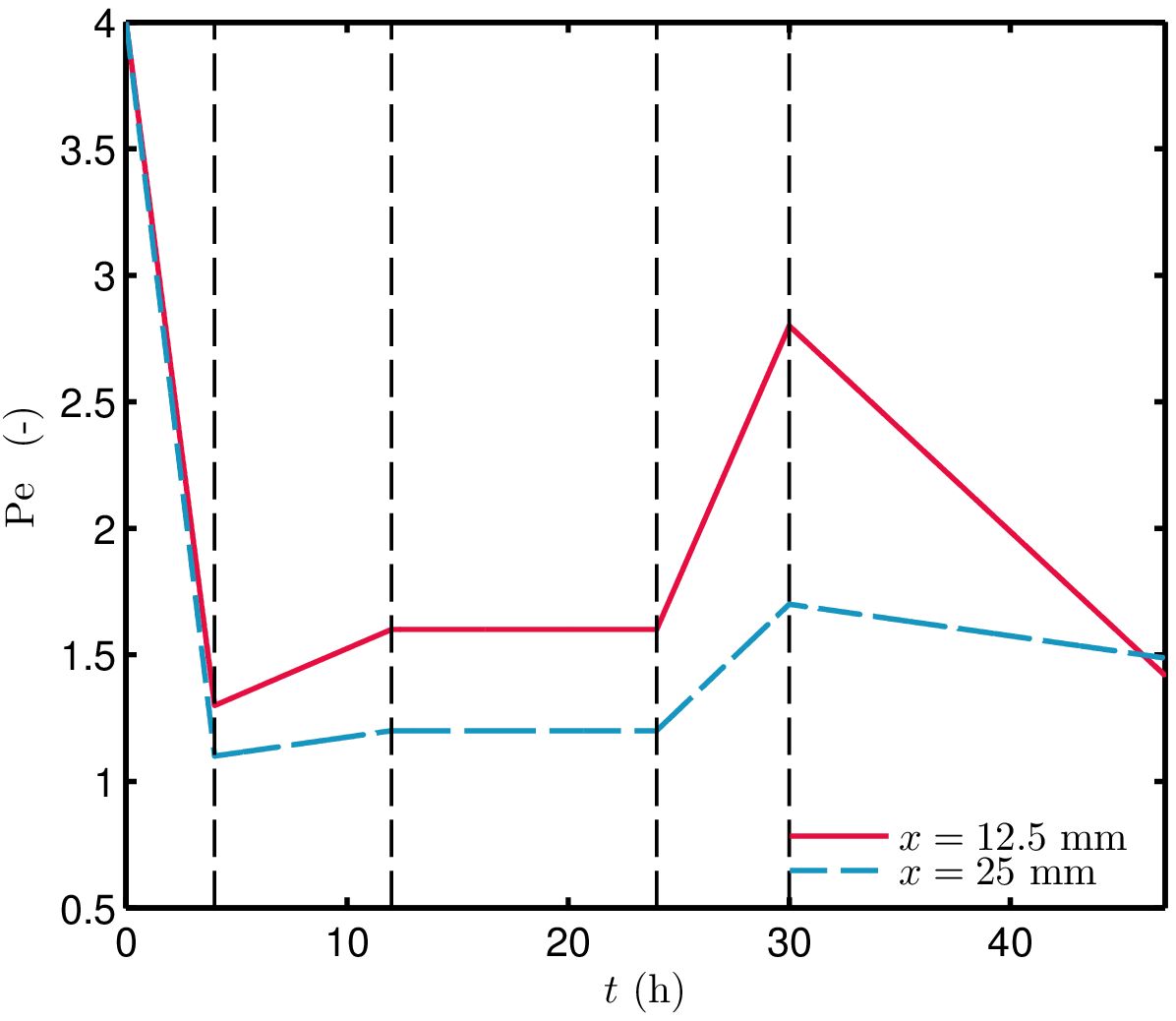}}
\caption{\small\em Time evolution of $\dfrac{1}{T}$ (a) and the estimated \textsc{P\'eclet} number (b).}
\end{figure}

The \textsc{P\'eclet} number varies as a function of the inverse of the temperature. In Section~\ref{sec:Moisture_convection}, it was assumed the variation of temperature in the material as negligible. Here, this assumption is disconsidered as the experimental data from \cite{James2010} also provides the temperature evolution in the material, as shown in Figure~\ref{fig_exp:1_T_ft} represented as $\frac{1}{T}$. Five different periods can be observed, corresponding to increase or decrease steps. To improve the results, the \textsc{P\'eclet} number has been estimated as a function of time and according to those five different periods. This estimation yields to a $\mathcal{L}_2$ error of $\varepsilon \egal 1.1 \cdot 10^{\,-3}$. Figure~\ref{fig_exp:Pe_ft} illustrates the time evolution of $\Pe$ at the two measurement points. As it can be noticed in Figures~\ref{fig_exp:Rh_ft_x1}~and~\ref{fig_exp:Rh_ft_x2}, with the correction on \textsc{P\'eclet} number, the numerical results fit better with the experimental data. These results highlight that $\Pe$ also varies with $x$, probably due to the variation of the air pressure (and therefore the mass average velocity) in the material.

\section{Conclusions}

Numerous studies in the literature reported slower transient behaviour of moisture evolution, obtained by numerical models that consider only moisture diffusion through porous materials, when compared to experimental data. Although, the discrepancies might come from different reasons such as hysteresis, uncertainties on moisture storage and transport coefficients, and material anisotropy, the results presented in this paper reveals that the advective moisture transfer may play an important role on providing much more accurate results for relative humidity range within 30-70\% and for materials with a microstructure composed of larger pores. To solve the advective-diffusive problem, two numerical schemes have been proposed and their efficiencies have been compared for both linear and nonlinear cases. The \SG scheme has been compared to the extensively used \CN ~approach. The \SG ~scheme is based on an implicit-explicit discretisation of the equations. It has been proposed in $1969\,$, based on the solution of the \textsc{Poincar\'e}--\textsc{Steklov} operator to compute an analytical expression of the diffusion and advection fluxes at each mesh interface. Performance of both schemes were analysed for the two case studies.

The first case study considered a linear convective transfer through a porous material. The schemes solution were compared to a reference solution obtained using \textsc{Chebyshev} functions. Results have shown that both are first-order accurate in space $\O(\dx)$ and second-order in time $\O(\dt)^2$. The solution computed with \SG ~scheme is more accurate. The second case study focused on nonlinear transfer, with material properties dependent on the vapour pressure field. The extension of the \CN ~and \SG ~schemes were given specially to treat the nonlinearities of the problem. A \mCN ~was proposed in order to avoid sub-iterations at each time step of the algorithm. Both \mCN ~and \SG ~schemes were used to compute the solution of the problem with accuracy. Results have shown that the error is proportional to $\O(\dt)$. The \SG ~scheme has a CFL restriction stability but is more accurate than the \mCN ~approach. Moreover, the stability of the \CN ~scheme does not neccessary imply an accurate solution \cite{Patankar1980}. The choice of the time discretisation $\Delta t$ is an important issue to represent accurately the physical phenomenon. Both schemes have the same order of CPU time. The classical \CN ~approach has shown to be twice as slow. Furthermore, the CPU time of the \SG ~scheme can be reduced using an adaptive time step $\dt$, thanks to the CFL  stability condition scaling with $\dx$ when considering large spatial discretisation.

In the last section, the \SG ~scheme was used to compute the solution of a moisture convective problem from the literature with experimental data provided from \cite{James2010}. The purpose was to highlight the impact of the advection on the moisture transfer. The moisture diffusion model reveals the slower transient behaviour whereas the results of the convective moisture model shows good agreement with the experimental data. The \textsc{P\'eclet} number, quantifying the advection transfer, has been estimated. A local sensitivity analysis has shown the importance of the \textsc{P\'eclet} number in the numerical model. The dependency of the \textsc{P\'eclet} number to temperature has also been highlighted. Thus, further work should be focused on models combining heat, air and moisture transfer considering both diffusion and advection terms and using the \SG ~numerical scheme for a fast computation of an accurate solution. The importance of the advective term should also be further investigated and compared to the capacitive and diffusive terms, considering hysteresis, for both pendular and funicular states.


\section*{Acknowledgements}

The authors acknowledge the Brazilian Agencies CAPES of the Ministry of Education, the CNPQ of the Ministry of Science, Technology and Innovation, for the financial support. The authors also would like to acknowledge Dr. L.~Gosse (IAC--CNR ``Mauro Picone'', Italy) for his precious discussions on numerical matters. 


\appendix
\section*{Nomenclature}

\begin{tabular*}{0.7\textwidth}{@{\extracolsep{\fill}} | c  l l| }
\hline
\multicolumn{3}{|c|}{\emph{Latin letters}} \\
$d_{\,m}$ & moisture diffusion & $[\mathsf{s}]$ \\
$c_{\,m}$ & moisture storage capacity & $[\mathsf{kg/m^3/Pa}]$ \\
$g$ & liquid flux & $[\mathsf{kg/m^2/s}]$ \\
$h_{\,v}$ & vapour convective transfer coefficient & $[\mathsf{s/m}]$ \\
$k$ & permeability & $[\mathsf{s}]$ \\
$L$ & length & $[\mathsf{m}]$ \\
$\Pc$ & capillary pressure & $[\mathsf{Pa}]$ \\
$\Ps$ & saturation pressure & $[\mathsf{Pa}]$ \\
$\Pv$ & vapour pressure & $[\mathsf{Pa}]$ \\
$R_v$ & water gas constant & $[\mathsf{J/kg/K}]$\\
$T$ & temperature & $[\mathsf{K}]$ \\
$\mathsf{v}$ & mass average velocity & $[\mathsf{m/s}]$ \\
\multicolumn{3}{|c|}{\emph{Greek letters}} \\
$\phi$ & relative humidity & $[-]$ \\
$\rho$ & specific mass & $[\mathsf{kg/m^3}]$ \\
$\kappa$ & air permeability & $[\mathsf{m^2}]$ \\
$\mu$ & dynamic viscosity & $[\mathsf{Pa.s}]$ \\
\multicolumn{3}{|c|}{\emph{Dimensionless parameters}} \\
$a$ & advection coefficient & $[-]$ \\
$\mathrm{Bi}$ & \textsc{Biot} number & $[-]$ \\
$\cms$ & storage coefficient & $[-]$ \\
$\dms$ & permeability coefficient & $[-]$ \\
$J$ & flux & $[-]$ \\
$g^{\,\star}$ & liquid flow & $[-]$ \\
$\Pe$ & \textsc{Peclet} number & $[-]$ \\
$u$ & field & $[-]$ \\
$\nu$ & diffusion coefficient & $[-]$ \\
\hline
\end{tabular*}


\clearpage
\addcontentsline{toc}{section}{References}
\bibliographystyle{abbrv}
\bibliography{biblio}

\end{document}